\documentstyle[aps,eqsecnum,epsf,epsfig]{revtex}
\begin{document}

\baselineskip=16pt

\draft

\title{\hspace{11cm} Preprint WSU-NP-4-98\\
Sensitivity of HBT Interferometry to the Microscopic Dynamics of Freeze-out. } 
\author{ A.  Makhlin and  E. Surdutovich}
\address{Department of Physics and Astronomy, Wayne State University, 
Detroit, MI 48202}
\date{September 4, 1998}
\maketitle
\begin{abstract}

We study the HBT interferometry of ultra-relativistic nuclear collisions
using a freezeout model in which free pions emerge in the
course of the last  binary collisions in the hadron gas. We show that the HBT
correlators of both identical and non-identical pions change with respect
to the case of independent pion production. Practical consequences for the
design of the event generator with the built in Bose-Einstein correlations
are discussed. We argue that the scheme of inclusive measurement of the 
HBT correlation function does not require the symmetrization of the multi-pion
transition amplitudes (wave-functions).

\end{abstract}

\pacs{25.75.Gz, 12.38.Mh, 24.85.+p, 25.75.Ld}

\section{Introduction}
\label{sec:SNIn}

Pion interferometry is expected to provide important information about the
space-time picture of ultrarelativistic nuclear collisions.  It has already
proved to be a sensitive tool for the detection of the collective motion of the
matter created in the course of ultrarelativistic nuclear collisions
\cite{MS,AMS,NA35,NA44}.  The primary goal of this paper is to show that under
favorable circumstances, interferometry is capable of detecting the difference
between various mechanisms of freezeout which is the last transient phase
before the regime of free streaming of hadrons.   The effect we rely upon is
entirely due to real interactions at the freezeout stage, and its magnitude may
serve as a measure of these interactions.  We predict that the normalized
two-particle correlator  of {\em non-identical} pions (and even for pairs
like $\pi p$, $pn$, etc.) must differ from the reference unit value at any
difference $\Delta k$ of the pions momenta. The normalized correlator of
identical pions does not approach unity at large $\Delta k$ and slightly
exceeds value of 2 at $\Delta k=0$. (In the idealized model which is used in
this paper to {\em demonstrate} these effects, they are small.) The second goal
is to discuss the effect of the multiparticle final states on the one- and
two-pion inclusive spectra. We demonstrate that  the contribution of these
states depends on the microscopic dynamical mechanism of freezeout.  We
explicitly show that the effect  of the multi-particle final states is  limited
by the actual range of the  interactions at the freezeout stage. No particles,
which are causally or (and) dynamically disconnected from the two-pion
inclusive probe, can contribute this effect. Our general conclusion is that the
problem of the  analysis of HBT data cannot even be posed without reference to
an explicit dynamical model. Finally, our model may serve as a prototype of a
realistic dynamical mechanism of freezeout, provided that the preceding stage
of evolution is a hot gas of hadrons.  Our results can be used for the
simulation of truly quantum two-pion distributions with the input from the
event generators based on semi-classical dynamics.

The paper is organized as follows: In Sec.~\ref{sec:SN1},  the effect is
explained at the introductory level. We pose the problem of interferometry (in
its most rigorous form) as a problem of a quantum transition specified by the
observables and the initial data in Sec.~\ref{sec:SN2}. Sec.~\ref{sec:SN3}
deals with the formulation of the pion interferometry problem and the results
of calculations for the hydrodynamic theory of multiple production. A more
realistic model of freezeout, which incorporates some elements of hadron
kinetics, is considered in Sec.~\ref{sec:SN4}. The analytic answer for the
model-based calculations is obtained by the end of this section. In
Sec.~\ref{sec:SN5}, we discuss the effect of multi-pion final states on the
inclusive two-pion spectrum and speculate about physical phenomena that may
lead to the increase of the effect. A practical issue of the event generator,
with the built-in Bose-Einstein correlations, is discussed in
Sec.~\ref{sec:SN6}.

\section{Physical motivation}
\label{sec:SN1}

The simplest and most popular introductory explanation of the principles of
HBT interferometry is as follows: The amplitude to emit (``prepare'') the pion
at the point $x_N$ and to detect it with the momentum $k_1$ is
$a_N(k_1)e^{-ik_1x_N}$.\footnote{ We essentially base this on the Dirac 
definition of the wave function as a transition amplitude,
$\psi_{k}(x)\equiv\langle k|x\rangle =\psi^{*}_{x}(k)$, which allows for two
complementary  interpretations: (i) the particle is {\em prepared} with 
momentum $k$ and {\em detected} at the point $x$, or, equivalently, (ii) the
particle is {\em prepared} at the point $x$ and {\em detected} with the
momentum $k$. The same interpretation remains valid for the entire hierarchy of
the multi-particle wave functions, $\langle k_1,k_2|x_1,x_2\rangle$, etc.}  For
a system of two pions prepared at points $x_N$ and $x_M$, and detected with the
momenta  $k_1$ and $k_2$, the transition amplitude (wave function) is a
superposition of two indistinguishable amplitudes,
\begin{eqnarray}
{\cal A}_{NM}(k_1,k_2) = a_N(k_1)e^{-ik_1x_N}a_M(k_2)e^{-ik_2x_M} +
a_N(k_2)e^{-ik_2x_N}a_M(k_1)e^{-ik_1x_M}~,
\label{eq:E1.1}
\end{eqnarray}
and for a system of distributed pion sources, the two-particle inclusive
spectrum is
\begin{eqnarray}
  {dN^{(2)} \over d{\bbox k}_{1} d{\bbox k}_{2} } 
 =2~\sum_{N,M} \big\{ |a_N(k_1)a_M(k_2)|^2 +{\rm Re}~
\big[(a_N(k_1)a_M(k_2)a^*_N(k_2)a^*_M(k_1)e^{-i(k_1-k_2)(x_N-x_M)}\big]\big\}~.
\label{eq:E1.2}
\end{eqnarray}

This scheme carries an implicit assumption that the pions are truly
independently {\em created} in the state of free propagation.  Unfortunately, 
in the literature, this property is mostly taken for granted, though in  the
real world, it takes place  only in very special occasions. For example,
independence of pion production in nuclear collisions has been used to describe
Bose-Einstein correlations  at Bevalac energies when the excited nuclear matter
is dominated by nucleons \cite{GKW}.  In that case, the bremsstrahlung of
pions, that accompanies scattering of nucleons, was suggested to provide 
independent pion sources.

\begin{figure}[htb]
\begin{center}
\mbox{ 
\psfig{file=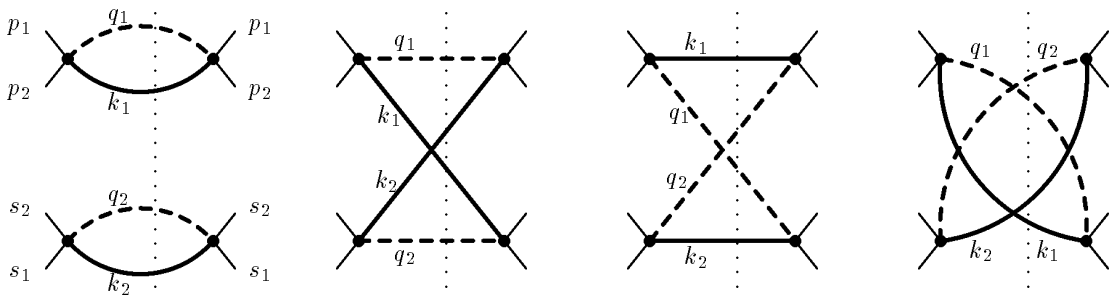,height=4cm,bb=103 548 548 665}
}
\end{center}
\caption{Diagrams for four terms in Eq.~(\ref{eq:E1.4}). Bold solid  and dashed
lines correspond to the detected and non-detected pions in the final state,
respectively. The second graph is absent when the detected pions are
non-identical. The vertical dotted lines cut the diagrams through the 
on-mass-shell lines corresponding to the final-state pions. The external legs
correspond to the initial-state pions.}
\label{fig:fig1}
\end{figure}

At RHIC energies (100 GeV/nucleon), we anticipate that the nuclear matter 
before the freezeout is totally dominated by pions and only very few nucleons
are expected in the central rapidity region. There is a long-standing
conjecture  that the last form of the matter, before the freezeout, is a hot
expanding hadronic gas. If this is indeed the case, then  the main type of
interaction at the freezeout stage is the binary collisions of  pions.
Therefore, the free propagation of any pion starts after it has experienced the
``last''  collision in the hadronic gas,  and there is no really independent
free pion production. Indeed, in this case, at least four particles are
involved in the interference  process, because at least two particles appear in
the final state in each binary collision. If two pions (say, $\pi^+\pi^+$) are
detected, then there are at least  two more undetected  particles emerging from
two different collisions, and these particles may be identical as well. Thus,
we encounter an additional (hidden) interference which  affects the measured
inclusive two-particle cross section.\footnote{ A similar mechanism is known to
provide corrections to the collision term in kinetic equations as well as the
first virial corrections to the equation of state of  quantum gases
\cite{Klimontovich}.}  Moreover, even if the detected particles are different
(e.g., $\pi^+\pi^-$), their partners still may be identical (e.g., the process 
$\pi^+\pi^0\to\pi^+\pi^0$ takes place at the coordinate $x_N$, and
$\pi^-\pi^0\to\pi^-\pi^0$ takes place at the coordinate $x_M$). Therefore, if
the last interaction is the binary collision in the pion gas, the na\"{\i}ve scheme
(\ref{eq:E1.1}) has to be modified.  In this modified case, there are four
interfering amplitudes,  
\begin{eqnarray}  
{\cal A}_{NM}(k_1,k_2;q_1,q_2) = 
a_N(k_1,q_1)a_M(k_2,q_2)e^{-i(k_1+q_1)x_N} e^{-i(k_2+q_2)x_M} \nonumber\\
+a_N(k_1,q_2)a_M(k_2,q_1)e^{-i(k_1+q_2)x_N} e^{-i(k_2+q_1)x_M}\nonumber\\
+a_N(k_2,q_1)a_M(k_1,q_2)e^{-i(k_2+q_1)x_N} e^{-i(k_1+q_2)x_M}\nonumber\\
+a_N(k_2,q_2)a_M(k_1,q_1)e^{-i(k_2+q_2)x_N} e^{-i(k_1+q_1)x_M}~.
\label{eq:E1.3}  
\end{eqnarray}  
The expression for the
two-particle spectrum thus becomes more  complicated,  
\begin{eqnarray} 
{dN^{(2)} \over d{\bbox k}_{1} d{\bbox k}_{2} }  
= \sum_{N,M}\sum_{q_1,q_2}|{\cal A}_{NM}(k_1,k_2;q_1,q_2)|^2 
=4 \sum_{N,M}\sum_{q_1,q_2}\big\{~|a_N(k_1,q_1)a_M(k_2,q_2)|^2\nonumber\\
+{\rm Re}~[~a_N(k_1,q_1)a_M(k_2,q_2)a^*_N(k_2,q_1)a^*_M(k_1,q_2)
e^{-i(k_1-k_2)(x_N-x_M)}\nonumber\\
+a_N(k_1,q_1)a_M(k_2,q_2)a^*_N(k_1,q_2)a^*_M(k_2,q_1)
e^{-i(q_1-q_2)(x_N-x_M)}\nonumber\\ +
a_N(k_1,q_1)a_M(k_2,q_2)a^*_N(k_2,q_2)a^*_M(k_1,q_1)
e^{-i(k_1+q_1-k_2-q_2)(x_N-x_M)}~]~\big\}~,  
\label{eq:E1.4}  
\end{eqnarray} 
and contains {\em three} interference terms (see Fig.~\ref{fig:fig1}). 
The first interference term 
corresponds to the familiar ``opened'' interference in the subsystem of the
detected pions. It is present only if the detected pions are identical. The two
other terms describe the ``hidden'' interference in the subsystem of undetected
pions which, however, are unavoidable partners of the detected pions in the
process of their creation in the final states. These terms are present even if
the detected pions are different. In fact, we deal with the {\em dynamically
generated} situation  when the system of two pions is  described by a density
matrix and not by the wave function.  When the dynamics of the system is driven
by binary collisions, the two pions just cannot be found in a pure
state!\footnote{At high pion multiplicity, we can neglect an exceptional case
when two detected pions come from the same binary collision.}  

A full scenario of ultrarelativistic heavy-ion collisions is still absent and
an understanding of the quantum kinetics from microscopic quantum field theory,
rather than from semi-classical approach, is only beginning to emerge.
Currently, the most self-consistent scenario is based on an assumption that
equilibration happens very early. Then it becomes possible to appeal to the
equation of state of nuclear matter at various stages of the scenario, starting
from the quark-gluon plasma (QGP) and ending up with the hadronic gas (as is
done, e.g., in Ref.\cite{Shuryak}.)  If this is not the case (no hadronic gas
occurs),  the pions may be created  in the  course of hadronization and
immediately  freely propagate. In this situation, it is more likely that the
pions are created independently. Thus, the difference between Eqs.
(\ref{eq:E1.1}), (\ref{eq:E1.2}) and (\ref{eq:E1.3}), (\ref{eq:E1.4}) becomes
of practical importance. Pion interferometry provides a tool which is capable
of distinguishing between  these two scenaria.  In Sec.~\ref{sec:SN5}, we argue
that the effect of hidden interference may become even stronger if the system
approaches the freezeout stage with the ``soft'' equation of state when  the
correlation length is long (as is, e.g., in the vicinity of a phase
transition).  

Furthermore, it is easy to understand that the correlations due to the hidden
interference  exist even, e.g., between pions and protons, protons and
neutrons, etc. All these effects, which may carry significant information
about the freeze-out dynamics, require (and deserve) special study, which is
beyond the scope of this paper.  We insist that it is highly desirable (though
not easy) either to measure all these correlation effects or to establish their
absence at a high level of confidence. Both results are important since they
would allow one to constrain the full self-consistent scenario of heavy-ion
collisions which is absent now.  These constraints may also affect theoretical
predictions of the photon and dilepton yields in heavy-ion collisions.

\section{Theoretical background: Definition of observables}
\label{sec:SN2}

A precise definition of observables is extremely important, because the HBT
interferometry does not allow one to pose the mathematically unambiguous
inverse problem. In general,  we have to formulate a model and compare the 
solution with the data, relying on common sense and physical intuition. The
full set of assumptions that accompany the formulation of the model is never
articulated in full. Particularly, the question about the nature of states in
which the particles are created is never discussed.  However, this is the key
issue. Interferometry is a consequence of the interference, and, in quantum
mechanics, the latter cannot be even addressed without direct  reference to the
quantum states. Indeed, interference takes place every time when, with a given
initial state, there are at least two alternative histories of evolution to a
given final state. The wave function is nothing but a transition amplitude
between the two states. HBT interferometry studies the two-particle wave
functions (or, more precisely, the two-particle density matrix). Therefore, the
most natural way to avoid any ambiguity is to pose the whole problem as a
problem of a quantum transition.\footnote{We closely follow (especially, in
Sec.~\ref{sec:SN4}) Ref.~\cite{phot}, where this approach was first used for
photon interferometry of the QGP. There, the photon emission was due to the
processes $q{\bar q}\to\gamma g$ and $qg\to\gamma q$, and the identical gluons
or quarks of the final state would interfere as well, if they had not
become the constituents of the thermalized system.}

Let $|{\rm in}\rangle$ be one of the possible initial states of the
system emitting a pion field which has three isospin components
\begin{eqnarray}
\mbox{\boldmath $\pi$}(x)\equiv\{\pi_i(x)\} =\{\pi_+(x),\pi_0(x),\pi_-(x)\}~,
\label{eq:E2.1}
\end{eqnarray}
where $\pi_-(x)=\pi_{+}^{\dag}(x)$,  and  $\pi_0(x)=\pi_{0}^{\dag}(x)$.
At the moment of the measurement ($t_f\to +\infty$) each component of the 
pion field can be decomposed into the system of analyzer eigenfunctions
$f_{\bbox k}(x)$,
\begin{eqnarray}
\pi_i(x) =\int d{\bbox k}[A_i({\bbox k})f_{\bbox k}(x)+
A^{\dag}_{i}({\bbox k})f^{\ast}_{\bbox k}(x)],~~~
f_{\bbox k}(x)=(2\pi)^{-3/2}(2k_{0})^{-1/2}e^{-i\,k\cdot x}~.
\label{eq:E2.3}
\end{eqnarray}
This expansion holds only after freezeout. In general, ${\bbox\pi}(x)$
is a multiplet of Heisenberg operators driven by the evolution operator
$S$. The annihilation operators $A_i({\bbox k})$  of the pion field are
given by
\begin{eqnarray}
A_{i}({\bbox k})\;=\;\int_{x^0=t_f} d^{3}x\; f^{*}_{\bbox k}(x)~i
\tensor{\partial_{x}^{0}}~\pi_i(x)~,~~~~
A_{i}^{\dag}({\bbox k})\;=\;\int_{x^0=t_f} d^{3}x\;\pi_{i}^{\dag}(x)~
i\tensor{\partial_{x}^{0}}~f_{\bbox k}(x)~.
 \label{eq:E2.4}
\end{eqnarray}
The operator $A_i({\bbox k})$ describes the effect of a detector (analyzer) 
far from the point of emission, so, by definition, the pion is detected on 
mass-shell, $k^0=({\bbox k}^2 +m^2)^{1/2}$.  The inclusive amplitudes to find 
one pion with
momentum ${\bbox k}$ and two pions with the momenta ${\bbox k}_1$  and 
${\bbox k}_2$ 
in the final state are
\begin{eqnarray}
\langle X| A_i ({\bbox k}) S |{\rm in}\rangle ~~~{\rm and}~~~
\langle X| A_i ({\bbox k}_1) A_j ({\bbox k}_2)S |{\rm in}\rangle~, 
\label{eq:E2.5}
\end{eqnarray}
respectively.  Here, the states $|X\rangle$ form a complete set of all possible
secondaries. Summing the squared moduli of these amplitudes over all
(undetected) states $|X\rangle$, and averaging over the initial
ensemble, we find the one-particle inclusive spectrum
\begin{eqnarray}
{ dN_{i}^{(1)} \over d{\bbox k} }  = {\rm Tr} {\hat \rho}_{\rm in}
\sum_{X} S^\dagger  A_{i}^{\dag}({\bbox k})|X\rangle\langle X|A_i({\bbox k})S
={\rm Tr} {\hat \rho}_{\rm in}
S^\dagger  A_{i}^{\dag} ({\bbox k})  A_i({\bbox k})S~,
\label{eq:E2.6}
\end{eqnarray}
where the density operator ${\hat \rho}_{\rm in}$ describes the
emitting system.  Of these two equations, the second one describes
the algorithm of the measurement.
In other words, we deal with the operator 
$~N_i({\bbox k}) =  A_{i}^{\dag}({\bbox k}) A_i ({\bbox k})$,
which gives the number of pions of the $i$'th kind detected by an
analyzer tuned to momentum ${\bbox k}$.  The first equation indicates
that all effects of multiparticle production are accounted for in this
{\em basic definition} of the inclusive one-particle spectrum. The
multiparticle states are encoded in a sum over the complete set of the
unobserved states,
\begin{eqnarray}
\sum_{X}|X\rangle\langle X|=1~,
\label{eq:E2.6a}
\end{eqnarray}
where they enter with an {\em a priori} assumption that they are
{\em identically  weighted}, and all together form
a unit operator. In the same way, one may obtain the inclusive two-pion 
spectrum
\begin{eqnarray}
  {dN^{(2)}_{ij} \over d{\bbox k}_{1} d{\bbox k}_{2} } 
= {\rm Tr} {\hat \rho}_{\rm in}
\sum_{X} S^\dagger  A_{i}^{\dag}({\bbox k}_1)A_{j}^{\dag}({\bbox k}_{2})
|X\rangle\langle X| A_{j}({\bbox k}_{2}) A_i({\bbox k}_1)S 
 = {\rm Tr} {\hat \rho}_{\rm in}\,  S^\dag
 A_{i}^{\dag}({\bbox k}_{1}) A_{j}^{\dag}({\bbox k}_{2})
 A_{j}({\bbox k}_{2})  A_{i}({\bbox k}_{1}) S~.
\label{eq:E2.7}
\end{eqnarray}
Once again, this equation defines the number of pairs, 
\begin{eqnarray}
 N_{ij}({\bbox k}_1,{\bbox k}_2)=
 A_{i}^{\dag}({\bbox k}_{1}) A_{j}^{\dag}({\bbox k}_{2})
 A_{j}({\bbox k}_{2})  A_{i}({\bbox k}_{1})=
 N_i({\bbox k}_1)[N_j({\bbox k}_2)-\delta_{ij}
 \delta({\bbox k}_1-{\bbox k}_2)]~,
 \label{eq:E2.8}
\end{eqnarray}
as the observable, and incorporates all multiparticle effects by its
derivation. This is a function of actual dynamics which is driven by the
evolution operator $S$ to select the states which physically contribute to the
process of the measurement, and to assign them the  {\em dynamically generated}
weights.

Equations (\ref{eq:E2.6}) and (\ref{eq:E2.7}) are universal in a sense that the
standard observables of interferometry are expressed in terms of their
Heisenberg operators. In order to make them useful, we have to specify both the
evolution operator $S$ and the initial data embodied in the density matrix
$\rho_{\rm in}$. In other words, the theory that claims to have any predictive
power  must  incorporate  physical information about the freezeout dynamics. 
Two models which reflect our current vision of the heavy-ion scenario, are
explored in the next two sections.

\section{Na\"{\i}ve freezeout.}
\label{sec:SN3}

The simplest and the most na\"{\i}ve model of freezeout has been explained in
detail in Ref.~\cite{MS}. In order to have a reference point for the discussion
of the  more involved situation detailed in the next section, we review this
model below with new updated emphases.  In fact, this model is not dynamical.
No microscopic mechanism of free pion production is specified. We just declare
the pion field to become free after some time (or, in a more intelligent way,
starting from some space-like surface $\Sigma_c$). This is what is done
technically, but certain assumptions must be kept in mind.   First, up to the
freezeout surface $\Sigma_c$, the system is assumed to be a continuous medium
which obeys relativistic hydrodynamic equations. The hypersurface $\Sigma_c$
corresponds to some critical temperature $T_c$ at which all interactions that
maintained the local thermal equilibrium in the expanding medium are switched
off. Consequently, we assume   the pion distribution over the momenta (not in
phase space!) at this moment to be (almost) thermal. It is incorporated into
the initial data for the future free propagation.

Second, the correlation length along $\Sigma_c$ is finite and much less than
the total size of the system which  is a dynamical consequence of the
interactions before the freezeout. In order to incorporate this property into
the na\"{\i}ve picture of the instantaneous freezeout, we must consider the whole
system as a collection of boxes (fluid elements) filled by free particles
which are opened when their world lines reach $\Sigma_c$.

Third, even though we may wish to disregard the quantum nature of the pions in
the fluid expansion phase,  we have to account for it when we pose  the
problem of interferometry. In other words, the pions have to be produced in
certain {\em states}.  This (perhaps, most important) goal is also achieved if
we mimic freezeout by the model of the opening boxes.

Practically, we act as follows: The field $\pi_i(x)$ in Eqs.(\ref{eq:E2.4})
has to be evolved in time starting from the initial data on the hypersurface 
$\Sigma_c$, i.e.,
\begin{eqnarray}
\pi_i(x)  =  \int d\Sigma_{\mu}(y)\:G_{\rm ret}(x-y)~
\tensor{\partial^{\mu}_{y}}~\pi_i(y)~.
 \label{eq:E3.1}
\end{eqnarray}
The space of states
in which the density matrix acts is also defined on this surface.
Substituting Eqs.~(\ref{eq:E3.1}) to (\ref{eq:E2.4}), we find 
the pion Fock operators, expressed in terms of the
initial fields,
\begin{eqnarray}
S^\dag  A_i({\bbox k})\, S  = 
\int d^{\,3}x\: f^{*}_{\bbox k}(x)~i\tensor{\partial_{x}^{\,0}}~
\int d\Sigma_{\mu}(y)\:G_{\rm ret}(x-y)~
\tensor{\partial^{\mu}_{y}}~ \pi_{i} (y)~.
\label{eq:E3.2}
\end{eqnarray}
This equation may be simplified using the explicit form
of the  free pion propagator,
\begin{eqnarray}
S^\dag  A_i({\bbox k}) S
 =  \theta(x^0-y^0)\int d\Sigma_{\mu}(y)
 f^*_{\bbox k}(y)~ i\tensor{\partial^{\mu}_{y}}~\pi_{i} (y)~.
\label{eq:E3.4}
\end{eqnarray}
The answer is simple because the whole problem of evolution is reduced to
just the free propagation of the pion field. Now, we have to incorporate the 
idea of the absence of long-range order on $\Sigma_c$. This is done in three 
steps. First, we replace the continuous integral
over $\Sigma_c$ by the sum of integrals over the ``cells'' labeled by a
discrete index $N$, 
\begin{eqnarray} \int_{\Sigma_c} d\Sigma (y) \to
\sum_{N}\int_{V^{\#}_N}d^3 y~, 
\label{eq:E3.5} \end{eqnarray} 
where $V^{\#}_N$ is the three-dimensional volume of the $N$-th cell 
on the hypersurface $\Sigma_c$.
(Hereafter, the $\#$-labeled quantities are related to the local reference 
frame with the time-axis normal to the hypersurface $\Sigma_c$.)
The pion field within the $N$-th cell also acquires an additional label
$N$. Next,  we perform the second quantization of the pion field within each
cell independently, 
\begin{eqnarray} 
\pi_N(y)=\sum_{\bbox p}[a_N({\bbox p})\phi_{\bbox p}(y)+a^{\dag}_{N}({\bbox p})
\phi^{\ast}_{\bbox p}(y)]~,\nonumber\\ \phi_{\bbox p}(y)=
(2~V^*_N p_{0})^{-1/2}e^{-i\,p\cdot x}~,~~~~~ p_{0}^{2}={\bbox p}^2+m^2~,
\label{eq:E3.6} 
\end{eqnarray} 
where the components of vector ${\bbox p}$ take discrete values defined by the
boundary conditions on the walls of each box, and the $\ast$-labeled quantities
are related to the local rest-frame of a fluid element. Finally, we impose the 
commutation relations on the pion field, 
\begin{eqnarray}
[a_N({\bbox p}_1),a^{\dag}_{M}({\bbox p}_2)]= 
\delta_{NM}\delta_{{\bbox p}_1{\bbox p}_2}~, 
\label{eq:E3.7} 
\end{eqnarray} 
which is equivalent to the dynamical independence of the fields belonging to
different cells. Each pion is created and propagates independently of all
others. Thus, the model is completely  defined. It is mathematically very
simple and reflects the main features of the physical process. There are
several scales in this model, the size $\ell$ of the elementary cell, the
freezeout temperature, $T_c$,  and the geometric parameters $L$ of the flow.
(The correlation length is of the order $\ell\agt 1/T_c$ and, in general, we
must require that $\ell \ll L$.)  The interplay of these parameters should be
explicitly accounted for in the course of calculations. The one-particle
spectrum of pions is 
\begin{eqnarray}
  {dN^{(1)} \over d{\bbox k} } 
 = \sum_{N}\sum_{\bbox p}\langle  a_N({\bbox p})a^{\dag}_{N}({\bbox p})\rangle
 {(k^{\#}_{0}+p^{\#}_{0})^2\over 4 V^{\ast}_N p^{\ast}_{0}}
 \int_{V^{\#}_N}d^3y e^{-i(k-p)y} \int_{V^{\#}_N}d^3y e^{+i(k-p)y}~.
\label{eq:E3.8}
\end{eqnarray}
At this point, we must treat one of the integrals as a delta-function which
sets the momenta ${\bbox p}^{\#}$ and ${\bbox k}^{\#}$ equal, while the second
integral becomes just the volume $V^{\#}_N$ of the fluid cell.  This procedure
requires that $|{\bbox k}|,~|{\bbox p}|~\gg~1/\ell\sim T_c~$, i.e., the
measured momenta should be sufficiently high. In the same way, we obtain the
expression for the two-particle spectrum. Since by virtue of
Eq.~(\ref{eq:E3.7}) we have
\begin{eqnarray}
\langle a_N({\bbox p}_1)a^{\dag}_{M}({\bbox p_2})
 a_{M'}({\bbox p'}_2)a^{\dag}_{N'}({\bbox p'_1})\rangle=\hspace{5cm}\nonumber\\ 
 = (\delta_{NN'} \delta_{MM'}
\delta_{{\bbox p}_1{\bbox p'}_1}\delta_{{\bbox p}_2{\bbox p'}_2}\!\!
+\delta_{NM'} \delta_{MN'}
\delta_{{\bbox p}_1{\bbox p'}_2}\delta_{{\bbox p}_2{\bbox p'}_1})
\langle a_N({\bbox p}_1)a^{\dag}_{N}({\bbox p_1})\rangle 
\langle a_M({\bbox p}_2)a^{\dag}_{M}({\bbox p_2})\rangle ,\nonumber
\end{eqnarray}
the two-pion inclusive spectrum becomes,
\begin{eqnarray}
  {dN^{(2)}_{ij} \over d{\bbox k}_{1} d{\bbox k}_{2} } =
  {dN^{(1)} \over d{\bbox k}_1 } {dN^{(1)} \over d{\bbox k}_2 } 
+ \sum_{NM}\sum_{\bbox p_1,p_2}
 \langle a_N({\bbox p}_1)a^{\dag}_{N}({\bbox p_1})\rangle 
\langle a_M({\bbox p}_2)a^{\dag}_{M}({\bbox p_2})\rangle \nonumber\\
\times {(k^{0\#}_{1}+p^{0\#}_{1}) (k^{0\#}_{1}+p^{0\#}_{2})
(k^{0\#}_{2}+p^{0\#}_{2}) (k^{0\#}_{2}+p^{0\#}_{1})
\over 4^2 V^{\ast}_N  V^{\ast}_M p^{0\ast}_{1} p^{0\ast}_{2}}
\cos(k_1-k_2)(x_N-x_M)  \nonumber  \\
\times \int_{V^{\#}_N}d^3y e^{-i(k_1-p_1)y} 
\int_{V^{\#}_N}d^3y e^{+i(k_1-p_2)y}
\int_{V^{\#}_M}d^3y e^{-i(k_2-p_2)y} 
\int_{V^{\#}_M}d^3y e^{+i(k_2-p_1)y}~. 
\label{eq:E3.10}
\end{eqnarray}
Thus, we have reproduced Eq.~(\ref{eq:E1.2}) which followed from an intuitive
conjecture (\ref{eq:E1.1}) about the interference of two indistinguishable
amplitudes. These amplitudes, by their design, reflect all relevant properties
of the interaction which prepare the two-pion system and a device which detects
this system. This {\em single device} consists of two detectors  (e.g., two
tracks) tuned to the pions with momenta ${\bbox k_1}$ and ${\bbox k_2}$.
Since, by the definition of the inclusive measurement, nothing else is
measured, there are two and only two interfering amplitudes. Regardless of how
large the total number, ${\cal N}$, of pions produced in a particular event is,
only the two-pion transition amplitude has to be symmetrized since the
remaining ${\cal N}-2$  pions are not measured.

In order to avoid any misunderstanding, we remind the reader that the
quantum-mechanical measurement of some observable, by definition, includes
the procedure of averaging over an ensemble. A single element of this ensemble
carries no quantum-mechanical information. Following Ref.\cite{TEV}, we can
argue that the inclusive measurement explores all quantum mechanical
fluctuations which can dynamically develop before the moment of measurement and
are consistent with the detector response. From this point of view, nothing but
a  causal chain of real interactions can affect the detector response, and only
dynamical histories can interfere. There is a significant difference between
the cases when one-, two-, or three-particle inclusive distributions are
measured. Each of these measurements is unique in a sense that they are all
{\em mutually exclusive} even if they are obtained from the same ensemble of
multiparticle events. For example, if the three-pion distribution is measured
(i.e., the average over an ensemble is accomplished), then the integration over
the momentum of the third pion does not result in the two-pion inclusive
distribution~\cite{review}. This example reflects a qualitative difference
between classical and quantum distributions. The former are defined
immediately in terms of {\em probabilities} while the later are defined via
{\em transition amplitudes} with the probabilities playing the secondary role.

The same conclusions follow from the field-theory formalism we employ. The
interference emerges as a strict consequence of the commutation relations
(\ref{eq:E3.7}) which incorporate a distinctive property of the hydrodynamic
model to localize the freezeout point for each pion independently. We start from
the quantum operator (\ref{eq:E2.8}) of the measured observable and trace the
two-pion signal back to its origin. Eventually, we arrive at the square of the
symmetrized two-pion amplitude, once again, regardless of the total number 
${\cal N}$of pions in a particular event. The states with many pions are
completely accounted for in Eqs.~(\ref{eq:E2.5})  and (\ref{eq:E2.7})  and,
consequently, in (\ref{eq:E3.8}) and (\ref{eq:E3.10}). However, they can
contribute to the one- and two-pion observables only via real interactions
which are absent in our oversimplified model of a na\"{\i}ve freezeout.  A  more
realistic example with the interaction is considered in the next section.

In Eq.(\ref{eq:E3.10}), we seemingly encounter a problem. If, e.g., 
the first of the integrals is considered as the delta-function which sets
momenta ${\bbox p}^{\#}_{1}$ and ${\bbox k}^{\#}_{1}$ equal,
then the value of the second one is not obvious. It occurs that
the second integral over the volume $V^{\#}_N$ becomes
\begin{eqnarray}
\int_{V^{\#}_N}d^3y e^{+i(k_1-k_2)y}=V^{\#}_N.
\label{eq:E3.11}
\end{eqnarray}
To prove this, we need more physical information. The interferometry would
have been impossible if we could have traced 
each pion back to the coordinate of its
emission or, in other words, if we could have built an ``optical image'' of the
source. According to the Rayleigh criterion, the latter is possible only if 
$ |\bbox {k_1-k_2}|~L~ \agt 1~$.
Therefore, we may observe interference only  provided the  last inequality
does not hold. Thus we have
$ |\bbox {k_1-k_2}|~ \alt 1/L ~ \ll~ 1/\ell$, which proves (\ref{eq:E3.11}).

In the continuous limit, by 
introducing an auxiliary ``emission function''  $J(k_1,k_2)$, 
\begin{eqnarray}
J(k_1,k_2)\;=\;\int_{\Sigma_{c}} d\Sigma_{\mu}(x)\:
{k^{\mu}_{1}+ k^{\mu}_{2}\over{2}}
\: n(k_{1}\cdot u(x))\: e^{-i(k_{1}-k_{2})x}~,
\label{eq:E3.14}\end{eqnarray}              
the expressions for the one- and two-particle inclusive spectra
can be rewritten in a form which is convenient for 
numerical computations:
\begin{eqnarray}
k^{0}\,{{dN_{1}} \over {d{\vec k}} } \;=\; J(k,k)~ ~,
\label{eq:E3.15}\end{eqnarray} 
and
\begin{eqnarray}
k^{0}_{1} k^{0}_{2}\,{{dN_{2}} \over {d{\vec k}_{1} d{\vec k}_{2} }
}\; =\;   J(k_{1},k_{1})\, J(k_{2},k_{2})\:+\: {\rm Re}\bigg
[J(k_{1},k_{2})\,J(k_{2},k_{1})\bigg ]~ ~ ~,
\label{eq:E3.16}\end{eqnarray}    
respectively.

In Ref.~\cite{AMS}, these equations were used to study interferometry for
several types of one--dimensional flow. Here, we are interested only in the
case of the boost-invariant geometry which must be very close to the reality of
RHIC. Indeed, any scenario initiated with a strong Lorentz contraction (up to
0.1~fm!)  of the nuclei cannot possess a scale associated with the initial
state. Hence, both at classical and  quantum levels, the system must evolve
with the preserved boost-invariance. In terms of the hydrodynamic theory of
multiple production \cite{Landau}, the absence of scale in  initial data of the
relativistic hydrodynamic equations immediately leads to the Bjorken
self-similar solution \cite{Bjorken} as the only  possible solution. Recent
analysis of quantum fluctuations at the earliest stage of heavy-ion collisions
\cite{TEV}  indicates the same. The overlap of Lorentz-contracted nuclei
converts them into a system of modes of {\em expanding plasma} with the global
boost-invariant geometry. This is a single quantum transition and it is not
compatible with the picture of gradually developing parton cascade
\cite{Geiger}.

Theoretical analysis of two-pion correlations, in the picture of ``na\"{\i}ve 
freezeout,'' from the phase of expanding hot pion gas was done in
Refs.~\cite{MS,AMS}. One of the predictions of these studies,  the so-called
$m_t$-scaling of the pion and kaon correlators, was confirmed by the data of
the NA35 and NA44 collaborations \cite{NA35,NA44} obtained from Au-Au
collisions at SPS energies ($\sim$10~GeV/nucleon). These data carry two
important messages. First, even when the nuclei are Lorentz-contracted only up
to the size of 1~fm, this is almost enough to bring about the boost invariant
regime of collective flow. There is no doubts that this picture will be even
more pronounced at RHIC. The second, less trivial consequence of the  observed
$m_t$-scaling is that {\em  the freezeout is sharp}; if the creation of the
final-state pions were extended in (local) time, the $m_t$-scaling would vanish
\cite{MSW}. Thus, addressing the pion production at RHIC, we have every reason
to rely on two facts: (i) the hadronic matter before the freezeout forms a
collective system, and (ii) this system is in the state of the self-similar
boost-invariant expansion.

For the immediate goals of this study, an advantage of the intensive
longitudinal flow is that it provides a window in the phase-space of the two
pions where the effect of the hidden interference  (explained in
Sec.~\ref{sec:SN1}, and computed in more detail in Sec.~\ref{sec:SN4}) is most
visible. This window corresponds to the measurement of the transverse size of
the longitudinally-expanding pipe using pairs of pions with the same rapidity. 
In this way, we employ a distinctive feature of the hydrodynamic-type sources
to localize the emission spectrum. The required localization in rapidity is
achieved by choosing pions with large  transverse momenta ($p_t\agt 3T\sim
3m_\pi$). This has a very simple physical explanation.  If the emitting system
is kinetically equilibrated (or even sufficiently chaotized) then the mean
energy per  particle is limited from  above. Thus, in the rest-frame of a fluid
element, the particles with large transverse momenta can have, on average, only
small longitudinal momenta. In other words, such particles are effectively
frozen into the collective hydrodynamic motion in the laboratory
frame.\footnote{This is the physical origin of the so-called $m_t$-scaling and
the qualitative basis for the saddle-point calculations below.} Measuring there
longitudinal rapidity after freezeout, we are most likely to measure the
longitudinal rapidity of the fluid at the  emission site.

The parameters of the model with the Bjorken geometry are the critical
temperature, $T_{c}~ (\sim m_{\pi})$, and the space--like freeze--out
hypersurface, defined by $t^{2}-z^{2} = \tau^{2} = {\rm const}$. The
coordinates and the four-velocity of the fluid are parameterized as
$x^\mu=(\tau\cosh\eta,{\vec r},\tau\sinh\eta)$, and  $u^\mu
(x)=(\cosh\eta,{\vec 0},\sinh\eta)$, respectively. The rapidity, $\eta$, of a
fluid cell is restricted to $\pm Y$ in the center-of-mass frame. We assume an
axially-symmetric distribution of hot matter in a pipe with  area $S_{\bot}
= \pi R^{2}_{\bot}$. The particles are described by their momenta, $k^{\mu}_{i}
= (k^{0}_{i},{\vec k}_{i},k^{z}_{i}) \equiv (m_{i}\cosh\theta_{i}, {\vec
k}_{i}, m_{i}\sinh\theta_{i})$, where ${\vec k}_{i}$ is the two-dimensional
vector of the transverse momentum,  $\theta_{i}$ is the particle rapidity in
$z$-direction, and $m^2_i \equiv m_{\perp i}^{2} = m^{2}+ {\vec k}_{i}^{2}$ is
the transverse mass. Let us introduce $2\alpha = \theta_1-\theta_2$, $2\theta =
\theta_{1}+\theta_{2}$.  The one-particle distribution is expected to be close
to a thermal distribution of pions at temperature $T=T_c$. Since the $m_\bot$
values of interest are larger than $T_c$, we may take this distribution in a
Boltzmann form, and use the saddle-point method to estimate the integrals
(\ref{eq:E3.16}). For the one-particle distribution these two steps yield
\begin{eqnarray}
{{dN}\over{d\theta_1 d{\vec k}_1}} \approx \tau S_\bot m_1
 \int^{Y}_{-Y} d\eta \cosh(\theta_1-\eta)
 e^{-m_1\cosh(\theta_1-\eta)/T_c} \approx \tau S_{\bot}m_1
 \sqrt{2\pi T_{c}\over{m_{1}}} \: e^{-m_{1}/T_c}~.
\label{eq:E3.19} \end{eqnarray}
The general expression for $J(k_1,k_2)$ is
\begin{eqnarray}
J(k_1,k_2) =  {1\over 2} \int\!\!d{\vec r}_1 f(r_1)
e^{i({\vec k}_1-{\vec k}_2){\vec r}_1}\tau 
\int^{\eta }_{-\eta }\!\! d\eta \ \bigg [(m_1+m_2)
\cosh(\theta-\eta )\,\cosh\alpha -
 (m_{1}-m_{2})\sinh(\eta -\theta)\sinh\alpha \bigg] \nonumber\\ 
\times\exp \bigg \{ -{1\over{T_c}} \bigg [(m_1+iF(m_1-m_2))
  \cosh(\eta -\theta)\cosh\alpha -(m_1+iF(m_1+m_2))
\sinh(\eta -\theta)\sinh\alpha\bigg ] \bigg \}~,
\label{eq:E3.20}
\end{eqnarray}
where $F=\tau T_c$. Once we consider only those pions with large transverse 
momenta, the integral over the rapidity $\eta$ can (and should)  be 
computed in the saddle-point approximation. This yields
\begin{eqnarray}    
R(k_{1},k_{2}) = {dN_2\over d y_1 d{\vec p}_1
d y_2 d{\vec p}_2}/ \bigg [ {dN_1\over d y_1 d{\vec p}_1} 
{{dN_1}\over{d y_2 d{\vec p}_2}}\bigg ] -1  =  
{1\over 4}  \;  f(|{\vec q}_{\bot}|R_{\bot})
\; {{g(z)\,g(1/z)} \over {[h(z)\,h(1/z)]^{3/2}}}  \nonumber \\ 
\times \exp
\bigg \{  -{\mu \over {T_{c}}}\: \bigg [ h(z)\cos {H(z)\over{2}}
+h({1\over{z}})
\cos {{H(1/z)}\over{2}}-z-{1\over z}\bigg ]\bigg \} \nonumber \\ 
\times \cos\bigg \{ {\mu \over   {T_{c}}}\bigg [h(z)\sin {H(z)\over{2}}
+h({1\over{z}})
\sin {{H(1/z)}\over{2}} \bigg ]  + 
{3\over 4}\bigg [H(z)+H({1\over {z}})\bigg ]
+G(z)+G({1\over{z}}) \bigg \}~,  
\label{eq:E3.21} 
\end{eqnarray} 
where $\mu =(m_{1}m_{2})^{1/2}$ and $z = (m_{1}/m_{2})^{1/2} $, 
and we have introduced the functions 
\begin{eqnarray}    
h(z) = \bigg \{ \bigg [
z^{2}-F^{2}(z-{1\over{z}})^{2}+4F^2\sinh^{2}\alpha \bigg ]^2 
+4F^2(z^2-\cosh 2\alpha)^2 \bigg \}^{1/4};\nonumber\\ 
g(z) = \bigg [(z^{2}+\cosh 2\alpha)^{2}
+F^{2}(z^{2}-{1\over{z^{2}}})^{2}\bigg ]^{1/2};\nonumber\\ 
\tan H(z) =
{2F(\cosh 2\alpha-z^{2}) \over {z^{2}-F^{2}(z-{1\over
{z}})^{2}+4F^{2}\sinh^{2}\alpha}};~~~~~ 
\tan G(z) = {F(z^{2}- 1/z^{2}) \over
{z^2+\cosh 2\alpha}}~. 
\label{eq:E3.22} 
\end{eqnarray}
When the transverse momenta of two pions are equal, then $m_1=m_2$, and 
Eq.~(\ref{eq:E3.21}) reproduces the result of Ref.~\cite{AMS}. The function $f$
depends on how we parameterize the transverse distribution of the matter in the
longitudinally expanding pipe.  If $f(r) = \theta (R^{2}_{\bot}-r^{2})$ then 
$~f(|{\vec \Delta k}_{\bot}|R_{\bot})= [2 J_1(|\Delta {\vec
k}|R_{\bot})/|\Delta{\vec k}|R_{\bot}]^2~$, where $J_1$ is a Bessel function.
This HBT correlator possesses a natural property of any two-point function in
the boost-invariant geometry; it depends on the difference of rapidities and
not on the difference of the longitudinal momenta. The $m_t$-scaling of the
``visible longitudinal size'' is just a synonym for this property, which
supports a picture of the sharp freeze-out of matter in the state dominated by
the boost-invariant-like longitudinal expansion. 

Two remarks are in order: 

1. The shape of the correlator given by Eqs.~(\ref{eq:E3.21}) and
(\ref{eq:E3.22}) is manifestly non-Gaussian. Thus, it is plausible to get rid
of the intermediate Gaussian fit  and to use  these equations to fit the data
as the first step.   

2. A main deficiency of the correlator (\ref{eq:E3.21}) is that the possible
transverse expansion has not been taken into account. This is not easy to do
since the initial data in the transverse plane are, as yet, poorly understood
and the amount of matter involved in the collective transverse motion  at the
freezeout stage may vary depending on the initial conditions, equation of
state, etc.  We consider this issue still open.

\section{Dynamical freezeout.}
\label{sec:SN4}

Now, let us formulate a more realistic model for the freezeout, relying on the
following phenomenological input.  Let the system before the collision be a
pion-dominated hadronic gas. Therefore, since pions are the lightest hadrons,
the regime of continuous medium is supported mainly due to $\pi\pi$-scattering.
In the framework of chiral perturbation theory there are two main channels
contributing to this process.  The first channel is  $s$-wave scattering which
we shall  model using the sigma-model. The second channel is the $p$-wave
$\pi\pi$-scattering via the $\rho$-meson.  We shall use a model interaction
Lagrangian of the form
\begin{eqnarray} 
{\cal L}_{\rm int}= {g_{4\pi}\over 4}(\bbox{\pi^\dag\cdot\pi})^2+
g_\sigma \sigma (\bbox{\pi^\dag\cdot\pi})+
{f_{\rho\pi\pi}\over 2} \bbox{\rho}^\mu {\bbox\cdot}[\partial_\mu 
\bbox{\pi^\dag\times\pi} + \bbox{\pi^\dag\times}\partial_\mu \bbox{\pi}]+
\cdot\cdot\cdot~,
\label{eq:E4.1} 
\end{eqnarray}
where $\sigma (x)$ is the field of the scalar $\sigma$-mesons, and 
${\bbox\rho}^\mu=(\rho^{-},\rho^{0},\rho^{+})$ is the field of the 
$\rho$-mesons.
The reader can easily recognize here that part of the Lagrangian of the
sigma-model which is necessary to reproduce the tree-level (or even skeleton)
amplitudes of the $s$- and  $p$-wave $\pi\pi$-scattering.   The evolution
operator is usually defined as
\begin{eqnarray}
S=T \exp\{i\int {\cal L}_{\rm int}(x)d^4x\}~.
\label{eq:E4.2} 
\end{eqnarray}
Applying the standard commutation formulae \cite{Bogol},
\begin{eqnarray}
A_{[i]}({\bbox k}) S - S A_{[i]}({\bbox k}) =\int d^4 x 
{\delta S \over \delta \pi^{\dag}_{i}(x)} f^{*}_{\bbox k}(x)~,~~~
S^{\dag} A^{\dag}_{[i]}({\bbox k})- A^{\dag}_{[i]}({\bbox k})S^{\dag}=
\int d^4 x
f_{\bbox k}(x){\delta S^{\dag} \over \delta \pi_{[i]}(x)}~,
\label{eq:E4.3} 
\end{eqnarray}
to the Eqs.~(\ref{eq:E2.6}) and (\ref{eq:E2.7}) we obtain
\begin{eqnarray}
{ dN_{[a]}^{(1)} \over d{\bbox k} }  = \int d^4y_1d^4y_2 f_{\bbox k}(y_1)
\bigg\langle {\delta S^{\dag}\over \delta\pi_{[a]}(y_1)}
{\delta S\over \delta\pi_{[a]}^{\dag}(y_2)}\bigg\rangle f^{*}_{\bbox k}(y_2)~,
\label{eq:E4.4}
\end{eqnarray}
for the one-particle spectrum of the pions of kind $a$ 
$(\pi^{+},\pi^{0},\pi^{-})$, and
\begin{eqnarray}
  {dN^{(2)}_{[ab]} \over d{\bbox k}_{1} d{\bbox k}_{2} } 
 = \int d^4y_1 d^4y_2 d^4y_3 d^4y_4 f_{{\bbox k}_1}(y_1)f_{{\bbox k}_2}(y_3)
\bigg\langle {\delta^2 S^{\dag} \over 
\delta\pi_{[a]}(y_1) \delta\pi_{[b]}(y_3)}
{\delta^2 S\over \delta\pi_{[a]}^{\dag}(y_2)  
\delta\pi_{[b]}^{\dag}(y_4)}\bigg\rangle  
f^{*}_{{\bbox k}_1}(y_2)f^{*}_{{\bbox k}_2}(y_4) ~,
\label{eq:E4.5}
\end{eqnarray}
for the spectrum for pairs of pions of kinds $a$ and $b$. According to our
agreement, the angular brackets denote an average weighted by the density
matrix $\rho_{\rm in}$. [In order to avoid confusion, we place the isospin
indices in square brackets.]

Since the goal of this paper is to demonstrate a physical effect, we shall
limit ourselves, in what follows, with the ${\bbox \pi}^4$ interaction term. In
this case, we are able to do most calculations analytically, which is necessary
in order to understand the interplay of the many parameters. More involved
calculations will require a realistic description of the $\pi\pi$-scattering
and Monte-Carlo computation of multiple integrals. 

Calculation of the one-pion spectrum is relatively simple. A direct computation
of the functional derivatives  in Eq.(\ref{eq:E4.4}) leads to 
\begin{eqnarray}
{ dN_{[a]}^{(1)} \over d\bbox k }  = {g_{4\pi}^2\over 4}\int d^4x d^4y
{e^{-ik(x-y)}\over 2k^0 (2\pi)^3}\sum_{i,j} \langle 
T^\dag(S^\dag:\pi^{\dag[i]}(x)\pi^{[i]}(x)\pi^{\dag[a]}(x):)
T(:\pi^{[a]}(y)\pi^{\dag[j]}(y)\pi^{[j]}(y):S)\rangle~,
\label{eq:E4.6}
\end{eqnarray}
where the symbols $T$ and $T^\dag$ denote the time and anti-time
orderings, respectively. 
Coupling the field operators according to the Wick theorem (to the lowest order
of the perturbation expansion we have to put $S=1$), the average in
Eq.~(\ref{eq:E4.6}) becomes,
\begin{eqnarray}
 \langle ~\cdot\cdot\cdot~\rangle=
 i^3[G^{[aa]}_{01}(y,x) G^{[ij]}_{10}(x,y) G^{[ji]}_{01}(y,x)+
 G^{[ja]}_{01}(y,x) G^{[ij]}_{10}(x,y) G^{[ai]}_{01}(y,x)]~,
\label{eq:E4.7}
\end{eqnarray}
where any pair of arguments $x$ and $y$ lie within the same cell of a size
defined by the correlation length in the system under consideration. This
is a consequence of the cellular structure of the density operator 
${\hat\rho_{\rm in}}$
which determines the average $\langle\cdot\cdot\cdot\rangle$.
The correlators $G_{AB}(x,y)$ are defined (using the Keldysh
technique \cite{Keld}; see also Ref.~\cite{TEV}) as
\begin{eqnarray}
 G^{[ij]}_{AB}(x,y)=
 -i\langle T_c (\pi^{[i]}(x_A)\pi^{\dag[j]}(y_B)\rangle~. 
\label{eq:E4.7a}
\end{eqnarray}
Taking for $a=1$ $(\pi^+)$ and using the facts that 
$G^{[ij]}\propto\delta^{ij}$, $~G^{[33]}_{01}(y,x)=G^{[11]}_{10}(x,y)$, and
$~G^{[33]}_{10}(y,x)=G^{[11]}_{01}(x,y)~$ we arrive at
\begin{eqnarray}
 \langle ~\cdot\cdot\cdot~\rangle=
 i^3 G^{[11]}_{01}(y,x) [3 G^{[11]}_{10}(x,y) G^{[11]}_{01}(y,x)+
  G^{[00]}_{10}(x,y) G^{[00]}_{01}(y,x)]~.
\label{eq:E4.8}
\end{eqnarray}
Next, we have to insert this expression into Eq.~(\ref{eq:E4.6}) and pass over
to the momentum representation (cell-by-cell) according to 
\begin{eqnarray}
 G^{[11]}_{01}(p)= G^{[33]}_{10}(-p)=
 -2\pi i[\theta(p^0)n^{(+)}(p) +\theta(-p^0)]~,\nonumber\\
G^{[11]}_{10}(p)=G^{[33]}_{01}(-p)=
 -2\pi i [\theta(p^0) +\theta(-p^0)n^{(-)}(p)]~, \nonumber\\
 G^{[00]}_{01}(p)=
 -2\pi i[\theta(p^0)n^{(0)}(p) +\theta(-p^0)]~,\nonumber\\
G^{[00]}_{10}(p)=
 -2\pi i [\theta(p^0) +\theta(-p^0)n^{(0)}(-p)]~.
  \label{eq:E4.9}
\end{eqnarray}

These equations reflect a very important feature of the process under
investigation (already accounted for in Eqs.~(\ref{eq:E4.4}) and 
(\ref{eq:E4.5}). )  Namely, the initial states in the system of colliding pions
are populated with the densities  $n_N^{(\pm)}(p)$ of charged pions and
$n_N^{(0)}(p)$ of  neutral pions. All the final states of free propagation {\em
are not occupied}. They are virtually present with an {\em a priori} unit
weight in the expansion (\ref{eq:E2.6a}) of the  unit operator. Access to the
states with many pions is provided dynamically by real interactions, and these
states show up in higher orders of the evolution operator expansion. Depending
on the type of commutation relations, these states always appear in a properly
symmetrized form, and there is no need to start with any {\em ad hoc} 
symmetrization.  By definition, the wave function is the transition amplitude
between the prepared initial state and detected final state. Unless the
transition is explicitly measured, the wave function does not exist at all and
there is no object for symmetrization.  Therefore, there is no statistical
correlations associated with the multiple production of pions. All 
correlations are dynamical (see further discussion in Sec.~\ref{sec:SN5}).

After some manipulations, we obtain for the one-particle spectrum,
\begin{eqnarray}
{ dN_{\pi^{+}}^{(1)} \over d\bbox k }={g_{4\pi}^2\over 4}\sum_{N} V^{(4)}_{N}
\int {d^4p_1 d^4p_2 d^4q\over 2k^0(2\pi)^{8} }\delta(k+q-p_1-p_2)
\delta(p_1^2-m^2) \delta(p_2^2-m^2) \delta(q^2-m^2)\nonumber\\
\times[3n^{(+)}_{N} (p_1)n^{(+)}_{N}(p_2) + 6n^{(+)}_{N}(p_1)n^{(-)}_{N}(p_2)+
2n^{(+)}_{N} (p_1)n^{(0)}_{N}(p_2))+n^{(0)}_{N}(p_1)n^{(0)}_{N}(p_2))]~.
\label{eq:E4.10}
\end{eqnarray}
Here, as in the model of the ``na\"{\i}ve freezeout'' discussed in
Sec.~\ref{sec:SN3}, one of the integrals over the space-time volume
$V^{(4)}_N$ of the domain (where the individual collision occurs), is treated
as the delta-function (viz., momentum conservation), while the second
integral becomes just the four-volume $V^{(4)}_{N}$ of the interaction
domain. (More details are given in Appendix A.) In what follows,
we shall limit ourselves to a simplified picture when all the statistical
weights of the initial state are taken in the Boltzmann form and the chemical
potential is zero. In this case, we gain an overall factor of 3+6+2+1=12 
and arrive at
\begin{eqnarray}
{ dN_{\pi^{+}}^{(1)} \over d\bbox k }  
= {12g_{4\pi}^2\over 4}\sum_{N} V^{(4)}_{N}
\int {d^4p_1 d^4p_2 d^4q\over 2k^0(2\pi)^{8} }\delta(k+q-p_1-p_2)
\delta(p_1^2-m^2) \delta(p_2^2-m^2)\delta(q^2-m^2)
e^{-\beta_N u_N(p_1+p_2)}~.
\label{eq:E4.11}
\end{eqnarray}
Now, the integration over $l=p_1-p_2$ can be explicitly carried out 
(see Eq.~(\ref{eq:A2.0})
\begin{eqnarray}
{ dN_{\pi^{+}}^{(1)} \over d{\bbox k }}  
= {12g_{4\pi}^2\over 4}\sum_{N} V^{(4)}_{N}
e^{-\beta_N (ku_N)}\int { d^4q\over 2k^0(2\pi)^{8} }{\pi\over 4}
e^{-\beta_N (qu_N)}\delta(q^2-m^2)\sqrt{1-{4m^2\over (k+q)^2}}~.
\label{eq:E4.12}
\end{eqnarray}
Introducing an auxiliary function,
\begin{eqnarray}
I(k,N,x_{NM})=\int d^4q\delta_+(q^2-m^2)e^{-\beta_N(qu)}~e^{iqx_{NM}}
\sqrt{1-{4m^2\over (k+q)^2}}~,
\label{eq:E4.12a}
\end{eqnarray}
which is computed in Appendix B, we may write explicitly,
\begin{eqnarray}
{ dN_{\pi^{+}}^{(1)} \over d{\bbox k }}  
= {12\pi g_{4\pi}^2\over 16}\sum_{N} {V^{(4)}_{N}\over 2k_0(2\pi)^8}
e^{-\beta_N (ku_N)}I(k,N,0) 
= {12g_{4\pi}^2\over 16}2\pi m^2{K_1({m / T})\over m/T}
\int {d^4x\over 2k_0(2\pi)^8} e^{-ku(x)/T}\nonumber\\
= {12g_{4\pi}^2\over 16(2\pi)^8}2\pi m^2{K_1({m / T})\over m/T}
\tau \Delta\tau S_\bot \sqrt{{2\pi T\over m_\bot}} e^{-m_\bot/T}~,
\label{eq:E4.12b}
\end{eqnarray}
where, with reference to the data discussed in Sec~\ref{sec:SN3},
we  assume that the freezeout occurs in a very small interval
$\Delta\tau$ of the proper time $\tau$. Additionally, we further simplify the
model by taking the temperature the same throughout the entire freezeout 
domain.

Calculation of the two-pion spectrum is very cumbersome but follows
exactly the same guideline,
\begin{eqnarray}
{ dN_{\pi^{a} \pi^{b}}^{(2)} \over d\bbox k_1 d\bbox k_2}  = 
{g_{4\pi}^4 \over 16}\int d^4y_1 d^4y_2 d^4y_3 d^4y_4
{e^{-ik_1(y_1-y_2)} e^{-ik_2(y_3-y_4)}
                   \over 4 k_1^0 k_2^0 (2\pi)^6}\nonumber\\
\times\sum_{i,j,l,n}
\langle T^\dag(S^\dag{\bbox :}\pi^{\dag [i]}(y_1)\pi^{[i]}(y_1)
\pi^{\dag [a]}(y_1){\bbox ::}
\pi^{\dag [l]}(y_3)\pi^{[l]}(y_3)\pi^{\dag [b]}(y_3){\bbox :})\nonumber\\
\times T({\bbox :}\pi^{[a]}(y_2)\pi^{\dag [j]}(y_2)~\pi^{[j]}(y_2){\bbox ::}
\pi^{[b]}(y_4)\pi^{\dag [n]}(y_4)\pi^{[n]}(y_4){\bbox :}S)\rangle~,
\label{eq:E4.13}
\end{eqnarray}
where, as before, we must put $S=1$ in the lowest order of the perturbation
expansion. If the observed pions are identical, then the result of coupling is
a very long expression which is given in its full form in Appendix A,
Eq.~(\ref{eq:A1.1}). The next step is to integrate over the elementary volumes
${\cal V}_{N}$. This is done  according to Eq.~(\ref{eq:A1.3}). Finally, using
the Boltzmann approximation for the statistical weights, we carry out an
explicit integration over $l_1=s_1-s_2$ and $l_2=p_1-p_2$ (see
Eq.~(\ref{eq:A2.0})).  The result is as follows,
\begin{eqnarray}
{ dN_{\pi^{+} \pi^{+}}^{(2)} \over d\bbox k_1 d\bbox k_2}  = 
{ dN_{\pi^{+}}^{(1)} \over d\bbox k_1 } { dN_{\pi^{+}}^{(1)} \over d\bbox k_2 }
~+~{\pi^2g_{4\pi}^4\over 16^2}\sum_{N,M}V^{(4)}_{N}V^{(4)}_{M} 
 e^{-\beta_N (k_1 u_N)}e^{-\beta_M (k_2 u_M)}\nonumber\\
\times \int { d^4q_1 d^4q_2  \over 4k_1^0 k_2^0(2\pi)^{16} }
\sqrt{1-{4m^2\over (k_1+q_1)^2}}\sqrt{1-{4m^2\over (k_2+q_2)^2}}
e^{-\beta_N (q_1 u_N)}e^{-\beta_M (q_2 u_M)}
 \delta_+(q_1^2-m^2) \delta_+(q_2^2-m^2) \nonumber\\
 \times {\rm Re}\{144~e^{-i(k_1-k_2)(x_N-x_M)}
+44 e^{-i(q_1-q_2)(x_N-x_M)}
+44e^{-i(q_1+k_1-q_2-k_2)(x_N-x_M)}\}~.
\label{eq:E4.14}
\end{eqnarray}

In this equation, we recognize the symmetrized transition amplitude for all
four pions engaged in the process of the two pion production. Each term
acquires a factor associated with the isospin algebra. Though all multi-pion
states are included in the definitions (\ref{eq:E4.4}) and (\ref{eq:E4.5}) of
the observables, it is neither necessary nor  even possible to symmetrize them,
if only one or two pions are measured. The next orders of the evolution
operator expansion will bring in additional final-state particles. They will
all be properly symmetrized. However,  these terms will be more and more
suppressed (discussion in Sec.~\ref{sec:SN5}).

In a similar way, the original expression, (\ref{eq:A1.2}),
 is transformed to the inclusive spectrum of two different pions,
\begin{eqnarray}
{ dN_{\pi^{+} \pi^{-}}^{(2)} \over d\bbox k_1 d\bbox k_2}  = 
{ dN_{\pi^{+}}^{(1)} \over d\bbox k_1 } { dN_{\pi^{-}}^{(1)} \over d\bbox k_2 }
~+~{g_{4\pi}^4\over 16^2}\sum_{N,M} V^{(4)}_{N}V^{(4)}_{M}
e^{-\beta_N (k_1 u_N)}e^{-\beta_M (k_2 u_M)}\nonumber\\
\times \int {d^4q_1 d^4q_2 \over 4k_1^0 k_2^0(2\pi)^{16} }
\sqrt{1-{4m^2\over (k_1+q_1)^2}}\sqrt{1-{4m^2\over (k_2+q_2)^2}}
 e^{-\beta_N (q_1 u_N)}e^{-\beta_M (q_2 u_M)}  
\delta_+(q_1^2-m^2) \delta_+(q_2^2-m^2)\nonumber\\
\times {\rm Re}[ 16~ e^{-i(q_1-q_2)(x_N-x_M)}+
26~e^{-i(q_1+k_1-q_2-k_2)(x_N-x_M)}]
\label{eq:E4.15}
\end{eqnarray} 
Next, we have to find the net yield of the interference terms after the 
momenta of the unobserved pions are integrated out. As a first step, we can use
an auxiliary function $I(k,x,\Delta x)$, (\ref{eq:E4.12a}) and rewrite
Eqs.~(\ref{eq:E4.14}) and (\ref{eq:E4.15}), replacing the discrete  sums by the
integration over the continuous distribution of the collision points,
\begin{eqnarray}
{ dN_{\pi^{+} \pi^{+}}^{(2)} \over d\bbox k_1 d\bbox k_2}  = 
{\pi^2g_{4\pi}^4\over 16^2}\int {d^4x_1 d^4x_1\over 4k_1^0 k_2^0(2\pi)^{16} }
 e^{-(k_1 u(x_1))/T_c}  e^{-(k_2 u(x_2))/T_c}\nonumber\\
\times {\rm Re}\{144 I(k_1,x_1,0) I(k_2,x_2,0)
[1+e^{-i(k_1-k_2)(x_1-x_2)}]\nonumber\\
+44I(k_1,x_1,(x_1-x_2))I^\ast(k_2,x_2,(x_1-x_2))
[1+e^{-i(k_1-k_2)(x_1-x_2)}]\}~,
\label{eq:E4.16}
\end{eqnarray}
for identical pions, and
\begin{eqnarray}
{ dN_{\pi^{+} \pi^{-}}^{(2)} \over d\bbox k_1 d\bbox k_2}  = 
{\pi^2g_{4\pi}^4\over 16^2}\int {d^4x_1 d^4x_1\over 4k_1^0 k_2^0(2\pi)^{16} }
 e^{-(k_1 u(x_1))/T_c}  e^{-(k_2 u(x_2))/T_c}\nonumber\\
\times {\rm Re}\{144 I(k_1,x_1,0) I(k_2,x_2,0)
+I(k_1,x_1,(x_1-x_2))I^\ast(k_2,x_2,(x_1-x_2))
[16+26 e^{-i(k_1-k_2)(x_1-x_2)}]\}~,
\label{eq:E4.17}
\end{eqnarray}
for  non-identical pions.  These equations are written in the same
approximation as the one-pion inclusive spectrum (\ref{eq:E4.12b}).

The first two terms of Eq.~(\ref{eq:E4.16}) have a well known form of the
product of the intensities of two independent sources and  an interference
function $[1+cos\Delta k \Delta x]$ and correspond to the standard scheme of
HBT interferometry. In these terms, the space-time integration over
coordinates $x_1$ and $x_2$ is factorized and reproduces  the result of the
na\"{\i}ve model of freezeout given by Eq.(\ref{eq:E3.21}). The only difference 
is an inessential kinematic factor and a form-factor 
\begin{eqnarray}
I(k_1,x_1,0) I(k_2,x_2,0)=
4\pi^2 m^4\bigg[ {K_1(m/T)\over m/T}\bigg]^2~,
\label{eq:E4.18}
\end{eqnarray}
which eventually cancels out in the normalized correlator.
In the last two terms of Eq.~(\ref{eq:E4.16}), we encounter an additional 
factor ${\cal F}(x_1,x_2)=I(k_1,x_1,\Delta x)I^\ast(k_2,x_2,\Delta x)$
which, in general, does not allow  one to factorize the space-time
integrations. In our special geometry of the freezeout, however, some further
simplifications are possible, provided the transverse momenta $k_{1t}$ and
$k_{2t}$ are large. Indeed, integrating with respect to $x_1$,
we end up with a modified version of the emission function (\ref{eq:E3.14}),
\begin{eqnarray}
J(k_1,k_2;x_2)=\int d^4 x_1
 e^{-k_{1}\cdot u(x_1))/T} e^{-i(k_{1}-k_{2})x_1}{\cal F}(x_1,x_2)\nonumber\\
=\Delta\tau\int d^2{\vec r_1}f({\vec r_1})
e^{i{\vec r_1}({\vec k_1}-{\vec k_2})}\int_{-Y}^{Y}\tau d\eta_1
e^{-(m_{1}/T)\cosh(\eta_1-\theta_1)-i\tau[m_{1}\cosh(\eta_1-\theta_1)-
m_{2}\cosh(\eta_1-\theta_2)]}{\cal F}(\Delta\eta,\Delta{\vec x})~,
\label{eq:E4.19}\end{eqnarray} 
where the form-factor ${\cal F}$ can be conveniently rewritten as
\begin{eqnarray}
{\cal F}(\Delta\eta,{\vec \rho})=
4\pi^2 m^4 \bigg| {K_1(m\sqrt{U^2})\over
              m\sqrt{U^2}}\bigg|^2~,
\label{eq:E4.20}
\end{eqnarray}
where              
\begin{eqnarray}
U^2={1\over T^2} +{\vec \rho}^{~2}+
4T\tau(T\tau+i)\sinh^2{\eta_1-\eta_2\over 2}~,\nonumber
\end{eqnarray}
and we denoted ${\vec \rho}={\vec r_1} -{\vec r_2}$. 
The saddle point of the integration over $\eta_1$ in Eq.~(\ref{eq:E4.19}) is
defined by the equation,\footnote{ The exponential in 
Eq.(\ref{eq:E4.19})  is the same as in
Eq.(\ref{eq:E3.20}). Thus, it is exactly the saddle point (\ref{eq:E4.21}),
which leads to  Eqs.~(\ref{eq:E3.21}) and (\ref{eq:E3.22}).}
\begin{eqnarray}
\tanh (\eta_1-\theta)={1+iT\tau[1+(m_2/m_1)] \over
1+iT\tau[1-(m_2/m_1)]}~\tanh\alpha~,
\label{eq:E4.21}
\end{eqnarray}
and is not affected by the form-factor (\ref{eq:E4.20}). Thus, if we select
pions with the same rapidity, then $2\alpha=\theta_1-\theta_2=0$ and the
saddle point occurs at the point $\eta_1=\theta=\theta_1=\theta_2$.
Integrating in the same way with respect to $x_2$, we come to the conclusion
that $\eta_2=\theta$. The form-factor then becomes independent of rapidities 
$\eta$, and its argument becomes a real function,\footnote{ When 
$\eta_1=\eta_2$, the form-factor ${\cal F}$ is a smoothly
decreasing function. When $\eta_1\neq\eta_2$, it acquires an additional 
oscillatory pattern in the $\Delta\eta$-direction.}
\begin{eqnarray}
{\cal F}(0,{\vec \rho})=
4\pi^2 m^4  {K_1^2(m\sqrt{(1/T^2)+{\vec \rho}^{~2}})\over
              m^2[(1/T^2)+{\vec \rho}^{~2}]}~.
\label{eq:E4.22}
\end{eqnarray}

\begin{figure}[htb]
\begin{center}
\mbox{ 
\psfig{file=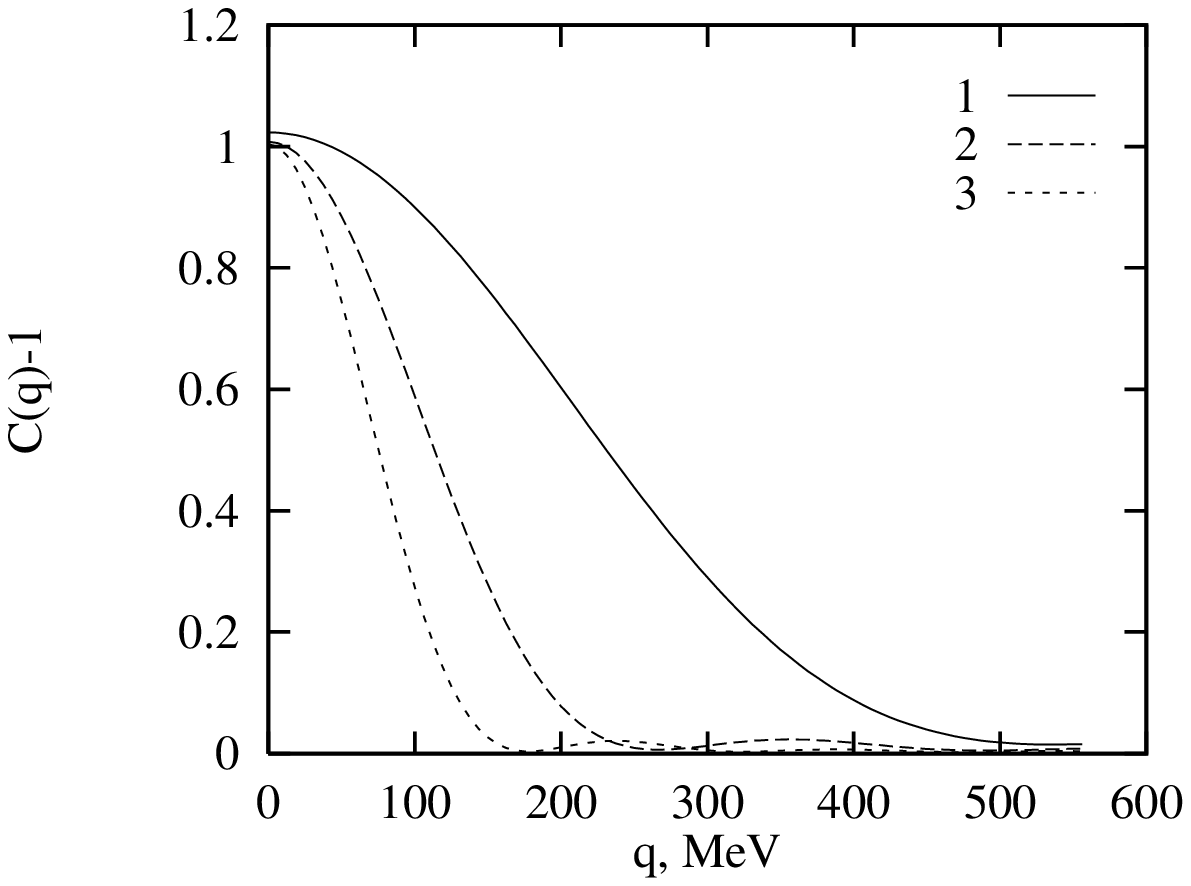,height=2.3in,bb=50 50 405 305}
\hspace{0.5cm}
\psfig{file=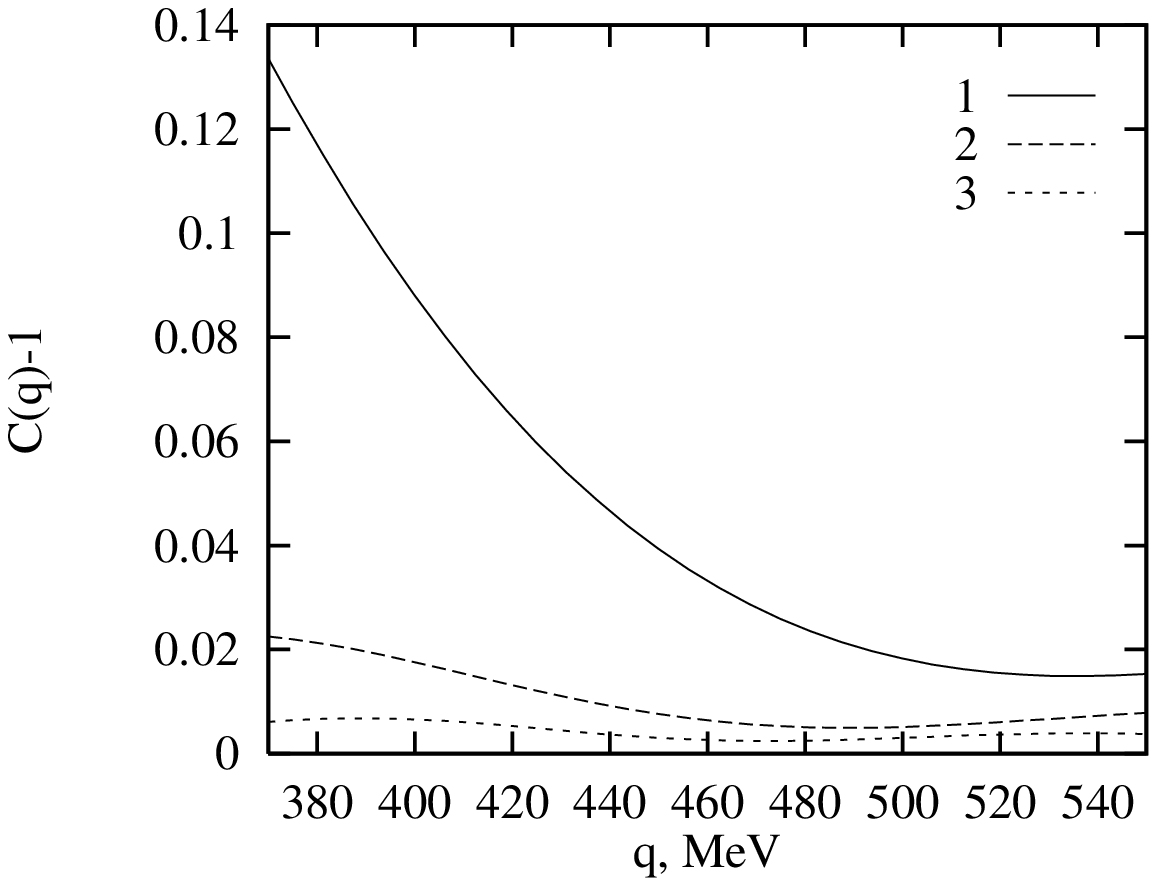,height=2.3in,bb=50 50 405 305}
}
\end{center}
\caption{Normalized correlator of identical pions as a function of  $q=(\Delta
k)_{\rm side}$ (with $(\Delta k)_{\rm out}=0$ and  $\Delta\theta =0$):  
(1) $mR_\bot=1$; (2) $mR_\bot=2$; (3)-- $mR_\bot=3$. The right plot is a
magnified portion of the left plot which allows one to see the scale of 
correction due 
to the hidden interference.}
\label{fig:fig2}
\end{figure}

\begin{figure}[htb]
\begin{center}
\mbox{ 
\psfig{file=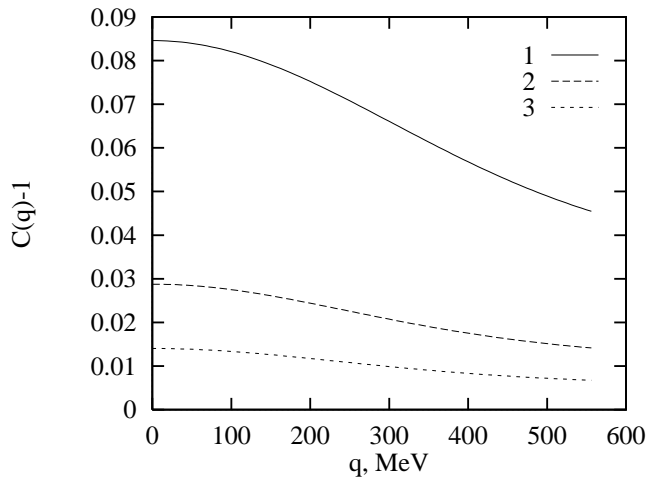,height=2.5in,bb=50 50 405 305}
}
\end{center}
\caption{Normalized correlator of non-identical pions. Notation is the same as
in Fig.~\ref{fig:fig2}}
\label{fig:fig3}
\end{figure}

From now on, we will deal only with  pions of the same
rapidity, which means that we physically take aim at two scattering processes
which occur in one  narrow slice of the longitudinally expanding pipe.
In this particular case, the very cumbersome formulae (like (\ref{eq:E3.21}))
become drastically simplified. 
The spectrum of two identical pions reads as
\begin{eqnarray}
{ dN_{\pi^{+} \pi^{+}}^{(2)} \over d\bbox k_1 d\bbox k_2}  = 
{ dN_{\pi^{+}}^{(1)} \over d\bbox k_1 }{ dN_{\pi^{+}}^{(1)}\over d\bbox k_2 }
~+~{\pi^2g_{4\pi}^4\over 16^2} {144 \over 4k_1^0 k_2^0(2\pi)^{16} }
4\pi^2 m^4 (\Delta\tau)^2 \tau^2{2\pi T\over \sqrt{m_1m_2}}
 e^{-(m_1+m_2)/T_c}  \nonumber\\
 \times\bigg\{{\rm Re}{\sqrt{m_1m_2}\over
 [m_1m_2+F(F-i)(m_1-m_2)^2]^{1/2}}
 \bigg[ \bigg(2\pi\int rdr f(r)J_0(|\Delta{\vec k}|r)\bigg)^2
 \bigg[ {K_1(m/T)\over m/T}\bigg]^2 \nonumber\\
+{44\over 144} 2\pi \int r_1dr_1 r_2dr_2 f(r_1)f(r_2)
\int_{0}^{2\pi}\!\!\!\! d\phi~
{K_1^2(m\sqrt{(1/T^2)+{\vec \rho}^{~2}})\over
 m^2[(1/T^2)+{\vec \rho}^{~2}]}J_0(|\Delta{\vec k}|\rho)\bigg] \nonumber\\
 +{44\over 144} 2\pi \int r_1dr_1 r_2dr_2 f(r_1)f(r_2)
 \int_{0}^{2\pi}\!\!\!\!d\phi~
{K_1^2(m\sqrt{(1/T^2)+{\vec \rho}^{~2}})\over
              m^2[(1/T^2)+{\vec \rho}^{~2}]}\bigg\}~,
\label{eq:E4.23}
\end{eqnarray}
where $\rho^2=r_1^2+r_2^2-2r_1r_2 \cos\phi$. In the same way, we obtain the 
two-particle spectrum of  non-identical pions,
\begin{eqnarray}
{ dN_{\pi^{+} \pi^{-}}^{(2)} \over d\bbox k_1 d\bbox k_2}  = 
{ dN_{\pi^{+}}^{(1)} \over d\bbox k_1 }{ dN_{\pi^{-}}^{(1)}\over d\bbox k_2 }
~+~{\pi^2g_{4\pi}^4\over 16^2} {144 \over 4k_1^0 k_2^0(2\pi)^{16} }
4\pi^2 m^4 (\Delta\tau)^2 \tau^2{2\pi T\over \sqrt{m_1m_2}}
 e^{-(m_1+m_2)/T_c}  \nonumber\\
 \times  2\pi \int r_1dr_1 r_2dr_2 f(r_1)f(r_2)
\int_{0}^{2\pi}\!\!\!\! d\phi~
{K_1^2(m\sqrt{(1/T^2)+{\vec \rho}^{~2}})\over
 m^2[(1/T^2)+{\vec \rho}^{~2}]}\nonumber\\
\times\bigg\{{16\over 144} + {26\over 144}~{\rm Re}\bigg[{\sqrt{m_1m_2}\over
 [m_1m_2+F(F-i)(m_1-m_2)^2]^{1/2}}\bigg]~J_0(|\Delta{\vec k}|\rho)\bigg\}~.
\label{eq:E4.24}
\end{eqnarray}
In order to give a quantitative estimate of the effect (yet with all
reservations intact due to the imperfection of our model description of
the $\pi\pi$-interaction) we shall assume that the distribution of the
longitudinally expanding hot pipe in the transverse direction is homogeneous
within a cylinder of radius $R_\bot$. We also neglect the transverse flow.
Next, we normalize the two-pion spectrum by the product of the one-pion
spectra and introduce
\begin{eqnarray}
C_{ab}({\bbox k_1},{\bbox k_2})={dN^{(2)}_{ab}/d{\bbox k_1} d{\bbox k_2}\over
(dN^{(1)}_{a}/d{\bbox k_1}) (dN^{(1)}_{b}/d{\bbox k_2})}~.
\label{eq:E4.25}
\end{eqnarray}
Finally, we consider only the case of $|{\bbox k_1}|=|{\bbox k_2}|$. 
Then, $m_1=m_2$ and we arrive at
\begin{eqnarray}
C_{\pi^{+}\pi^{+}}= 1+
\bigg({2J_1(|\Delta{\vec k}|R) \over |\Delta{\vec k}|R}\bigg)^2
+{11\over 36}\int_{0}^{R}{r_1dr_1 r_2dr_2\over (\pi R)^2}
\int_{0}^{2\pi}\!\!\!\!\! d\phi~ {2\pi\over K_1^2(m/T)}~
{K_1^2[(m/T)\sqrt{1+T^2{\vec \rho}^{~2}}]\over 1+T^2{\vec \rho}^{~2}}
[1+J_0(|\Delta{\vec k}|\rho)]~,
\label{eq:E4.26}
\end{eqnarray}
\begin{eqnarray}
C_{\pi^{+}\pi^{-}}= 1+
\int_{0}^{R}{r_1dr_1 r_2dr_2\over (\pi R)^2}
\int_{0}^{2\pi}\!\!\!\!\! d\phi~ {2\pi\over K_1^2(m/T)}~
{K_1^2[(m/T)\sqrt{1+T^2{\vec \rho}^{~2}}]\over 1+T^2{\vec \rho}^{~2}}
\bigg[{1\over 9}+{13\over 72}J_0(|\Delta{\vec k}|\rho)\bigg]~,
\label{eq:E4.27}
\end{eqnarray}
where we remind the reader that the longitudinal rapidities of the pions are 
set equal. The remaining integrations in Eqs. (\ref{eq:E4.26}) and
(\ref{eq:E4.27}) are performed numerically and the results are plotted in
Figs.~\ref{fig:fig2} and \ref{fig:fig3}.


\section{Possible contributions of the multiparticle states.}
\label{sec:SN5}

As we have seen in the previous section, the multi-pion final states indeed
contribute to the one- and two-pion inclusive cross sections. This contribution
is due to the real interactions, and therefore, we may speculate about its
magnitude in different situations and at different conditions of observation.
Within the unrealistic ${\bbox\pi}^4$-model we employed to make analytic
solution possible, the effect is understandably small. Indeed, the additional
pions emitted in the collision process lead to the form-factor
(\ref{eq:E4.22}),  ${\cal F}({\vec\rho})$, in the interference term. Let us
consider the  formal limit of $T\to\infty$. Then the form-factor becomes 
\begin{eqnarray} 
{\cal F}({\vec\rho},T\to\infty)= 
4\pi^2 m^4  {K_1^2(\;m\sqrt{{\vec\rho}^{~2}}\;)\over m^2{\vec\rho}^{~2}}~. 
\label{eq:E5.1} 
\end{eqnarray}
This expression is immediately recognized as the square of the pion vacuum 
correlator, 
\begin{eqnarray}
G_1({\vec\rho})=-i\langle 0|\pi ({\vec\rho} )\pi^\dag(0)+
\pi^\dag(0)\pi({\vec\rho})|0\rangle~, \nonumber
\end{eqnarray} 
which represents the density of states in the pion vacuum.  These are exactly
the states which are excited as the final states of free propagation in the
course of the $\pi\pi$-collision, and the characteristic scale of correlation
for these vacuum  fluctuations can be nothing but $m_{\pi}^{-1}$. At 
$T\to\infty$, the colliding pions are (formally) very hard (their thermal
wave-length becomes very short) and they cannot explore any scale except this
one. Thus, the ``hidden interference'' is active only if two collisions occur
within the range of a typical fluctuation in the ``trivial'' pion  vacuum.
Otherwise, these collisions are dynamically disconnected and the interference
becomes strongly suppressed.   At finite temperatures, the behavior of the
form-factor at the origin is less singular, but the main scale for the distance
$\rho$ remains   $m_\pi^{-1}$, which defines the smallness of the effect in the
oversimplified model ${\bbox\pi}^4$. The parameter which regulates the
magnitude of the excess of the {\em normalized} correlators (\ref{eq:E4.26})
and (\ref{eq:E4.27})  is the ratio of the effective radius of the dynamical
correlations to the full transverse size $R_\bot$ of the expanding pipe. As is
seen from the plot of Fig.~\ref{fig:fig3}, at $T=m_\pi$ the excess reaches
0.08  when   $R_\bot\simeq m_{\pi}^{-1}$. 

This understanding of the nature of the scale of correlation shows that the
magnitude of the effect may be quite different when the $\pi\pi$-interaction is
treated more adequately. For example, with properly fitted parameters, the
$\sigma$-model provides a good description of many data which depend on
$\pi\pi$-interaction.  Most of low-energy interactions of elementary particles
are mediated via wide resonances like $f_0$ (which is sometimes identified
with  $\sigma$), $a_1$, $\rho$, etc.  In this picture, the $\pi\pi$-interaction
is not as local as in the ${\bbox\pi}^4$-model, and more scales show up
resulting in the dependence of the $\pi\pi$-cross section on the momentum
transfer. The low-energy dynamics is rich and includes a chiral phase
transition. In the vicinity of this phase transition the correlation radii
should increase. Thus, on the one hand, we may speculate about various
dynamical phenomena which may lead to an increase of the correlation radii. On
the other hand, the system may develop a strong transverse hydrodynamic motion.
In this case, the ``measured $R_\bot$'' also becomes a dynamically defined
effective quantity, which is (though not as much as in the longitudinal
direction) smaller than the natural transverse size. 

Until now, we have considered the lowest order (with respect to the
interaction) contributions to the one- and two-pion inclusive distributions.
The next orders can be accounted for by the expansion of the evolution operator
in the Eqs.~(\ref{eq:E4.6}) and (\ref{eq:E4.13}). Even without explicit
calculations, it is clear that there will be two types of corrections. The
virtual radiative corrections should be absorbed into the definitions of the
dressed propagators and vertices. Within the framework of our phenomenological
approach, these are of no interest or value. The real corrections are connected
with the production of additional pions and involve integration over the
positions  of the additional points of emission (collision) and over the
momenta of the additional non-observed pions. The corresponding complicated
transition amplitudes will be properly symmetrized. However, the result of the
integration over the unobserved momenta will again be a set of form-factors
which will cut off all distances between the points of scattering which exceed
the correlation length. Thus, the higher order effects will always be
suppressed with respect to the lowest order.

Our last remark is about the possibility of the induced emission of pions. To
address this question, one has to realize that the so-called Bose-enhancement
has two different (and mutually exclusive) aspects. If we deal with the
measured states, then the symmetrization is solely due to the measurement. This
is the essence of the HBT effect. If the states are not measured, then we
indeed deal with the induced emission which is due to the  factor $\sqrt{n+1}$
in the equation $~a^\dag|n\rangle = \sqrt{n+1}|n+1\rangle~$, which requires
that the process of the emission of the $(n+1)$-th quantum physically occurs in
the field of all $n$ {\em identical} quanta emitted  before. This is the basic
idea behind the laser.  If the pions are produced  with space-like separation,
then stimulated emission is impossible.

\section{Bose-Einstein correlations in event generators}
\label{sec:SN6}

Our estimates show that the effect of hidden interference is not expected to be
large unless the system undergoes the freezeout starting from a soft mode with
long-range correlations.  This possibility is very appealing due to
its possible physical richness. Careful simulations will be required in
order to estimate the signal-to-noise ratio in a real detector. This job is
usually done with the aid of the so-called event generators. A complexity
stemming from an additional interference makes simulations more challenging. 
We address this issue below.

Over the last decade, many computer codes which are supposed to mimic  the
particle yields in heavy-ion collisions have been developed.  These codes rely
on classical propagation of the particles in hadronic cascades. They do not
describe the quantum properties of the measurement which lead to the
interference of amplitudes.  At best, the output of an event generator is the
list of  particles with one-to-one correspondence between their momenta and the
coordinates of their emission. Such a list is equivalent to a complete optical
image of the source. Therefore, there are no alternative histories in the free
propagation.  An ``afterburner'' has to be added to incorporate quantum
effects.

In order to recover the alternative histories and thus  simulate the two-pion
correlations in a model with independent pion production,  one has to know four
one-particle amplitudes, $a_N({\bbox k})$. They are necessary to recover the
two-pion amplitude ${\cal A}_{NM}(k_1,k_2)$ (see Eq.~(\ref{eq:E1.1})~).  Hence,
one has to store some information about the pion sources before the emission
which allows for such a reconstruction.  In the model of binary collisions,
the parameters of parents (their ID, momenta, and the coordinates of the
reaction volume) have to be stored for each pion from the detected pair. 
Otherwise, it is impossible  to reconstruct all four amplitudes which interfere
in the process of the  two-pion measurement. 

Eventually, one has to average the square of the inclusive amplitude over
an ensemble of all events.  Then, this procedure has to be repeated for every
point in the two-particle momentum space.  In more sophisticated models, with
the pions coming from decays of resonances, we may have even more interfering
amplitudes and each case requires special consideration \cite{RES}. 

The requirement that the quantum mechanics of the last interaction, which forms
the two-pion spectrum, must be recovered, follows from first principles and
cannot be removed. In some particular cases, one may take a short cut and
weaken the requirements of the complete description. However, every precaution
should be taken that the initial state of the interfering system is described
as a {\em quantum state}.

Finally, we have to mention that in the literature, another kind of correlation
function, which has been introduced \cite{Zaic} (see also
Refs.~\cite{Pratt,Wiedem}) in connection with the problem of simulation of
Bose-Einstein correlations in multi-pion events, is widely discussed. The
algorithmic definition of this object (which is different from the inclusive
differential distribution used in this paper) has several steps: (i) The events
are sorted by the total number ${\cal N}$ of pions in the event. (ii)  The 
correlation function $C_{\cal N}(k_1,...,k_{\cal N})$ is {\em measured},  i.e.,
the ${\cal N}$-pion amplitude is fully  symmetrized and its squared modulus is
averaged over the subset of events with this given ${\cal N}$. (iii) All but
two arguments of $C_{\cal N}(k_1,...,k_{\cal N})$ are integrated out, which
results in a set of two-pion correlators $C_{\cal N}(k_1,k_2)$, still dependent
on  ${\cal N}$. (iv)  This set is averaged over ${\cal N}$, in order to obtain
$C(k_1,k_2)$. We could not find a quantum observable which would correspond to
this correlator. We also could not establish a direct connection between this
correlator and the parameters of the emitting system.

\section{Conclusions}
\label{sec:SN7}

In this paper, we show that precise measurements of two-particle correlators,
both  for identical and non-identical particles,  may uncover important
information about the freezeout dynamics. A reliable theoretical prediction for
the magnitude of these correlations can be made only within the framework of an
elaborated scenario of the heavy-ion collision which is currently absent.  Data
may be very helpful for the theoretical design of the fully self-consistent
scenario.

It is equally important to measure the magnitude of the effect of hidden 
interference or to establish its absence at a high level of confidence. From
this point of view, our calculations carry an important message that there are
no  theorems which require the normalized two-pion correlator to  approach
unity for large differences of momenta. Neither should it take the value of 2 at
$\Delta k=0$. The latter value is the norm (charge) of the state with two
identical pions in a {\em pure} state only. In general, two final-state pions
are described by a density matrix. For the same reason, the normalized
correlator of non-identical pions should not equal 1. Moreover, the
correlator should not equal 1 even for the couples like $\pi p$, $pn$, etc.
Therefore, the technique of ``mixing of events'' should be used  with 
extreme caution, if used at all. The idea of a universal Gaussian fit of the HBT
correlators can hardly be fruitful as well. (The authors fully realize that there
may be many technical difficulties in the practical implementation of this
advice.) 

One more practical lesson of our analysis concerns the theory of multi-pion
correlations. This issue cannot even be addressed without an explicit account
of the dynamical mechanism responsible for the creation of multi-particle
states. Our general conclusion is that HBT interferometry cannot be
model-independent.

\vspace{1cm}

\noindent {\bf ACKNOWLEDGMENTS}

We are grateful to Rene Bellwied, Scott Pratt, Claude Pruneau, Edward Shuryak 
and Vladimir Zelevinsky for many stimulating discussions.  Conversations with
the members of the STAR collaboration helped us realize many  practical
problems of measurements in HBT.  We appreciate the help of Scott Payson for
critically reading the manuscript.

This work was supported by the U.S. Department of Energy under
Contract  No. DE--FG02--94ER40831.       

\bigskip

\renewcommand{\theequation}{A.\arabic{equation}}
\setcounter{equation}{0}

\section*{Appendix A.}
\label{sec:A1}

Here, we reproduce lengthy expressions for the two-particle inclusive spectra
as they appear after performing all couplings. We do not use them in this form
throughout the paper. However, they are the starting point, would we wish to
design a code for the simulation of the two-pion correlations. In this code,
the particle densities $n_N(p)$ will serve as the frequencies for the
initial-state pions to appear within the range of the last interaction. At
least some part of the  integrations with respect to the momenta of the
unobserved final-state pions is preferable to do analytically in order to
simplify the remaining Monte-Carlo integrations.

The full expression for the two-particle spectrum of identical pions as it 
appears after performing all couplings in Eq.(\ref{eq:E4.13}), is
\begin{eqnarray}
{ dN_{\pi^{+} \pi^{+}}^{(2)} \over d\bbox k_1 d\bbox k_2}  = 
{ dN_{\pi^{+}}^{(1)} \over d\bbox k_1 } { dN_{\pi^{+}}^{(1)} \over d\bbox k_2 }
~+~{g_{4\pi}^4\over 16}\sum_{N,M} 
\int {d^4s_1 d^4s_2 d^4p_1 d^4p_2 d^4q_1 d^4q_2 
   \over 4k_1^0 k_2^0(2\pi)^{24} }\nonumber\\
\times\delta_+(q_1^2-m^2) \delta_+(q_2^2-m^2)\delta_+(s_1^2-m^2) 
\delta_+(s_2^2-m^2) \delta_+(p_1^2-m^2) \delta_+(p_2^2-m^2)  \nonumber\\
\times \bigg\{[3n^{(+)}_{N} (s_1)n^{(+)}_{N}(s_2) +
6n^{(+)}_{N}(s_1)n^{(-)}_{N}(s_2)+
2n^{(+)}_{N}(s_1)n^{(0)}_{N}(s_2)+
n^{(0)}_{N}(s_1)n^{(0)}_{N}(s_2)]\nonumber\\
\times[3n^{(+)}_{M} (p_1)n^{(+)}_{M}(p_2) 
+ 6n^{(+)}_{M}(p_1)n^{(-)}_{M}(p_2)+
2n^{(+)}_{M}(p_1)n^{(0)}_{M}(p_2)+
n^{(0)}_{M}(p_1)n^{(0)}_{M}(p_2)]\nonumber\\
\times e^{-i(k_1-k_2)x_{NM}} 
\int_{{\cal V}_N}\!\! dy_1 e^{-i(k_1+q_1-s)y_1} 
\int_{{\cal V}_N}\!\! dy_2 e^{i(k_2+q_1-s)y_2}
\int_{{\cal V}_M}\!\! dy_3 e^{-i(k_2+q_2-p)y_3} 
\int_{{\cal V}_M}\!\! dy_4 e^{i(k_1+q_2-p)y_4}\nonumber\\
+[26 n^{(+)}_{N} (s_1)n^{(-)}_{N}(s_2) n^{(+)}_{M} (p_1)n^{(-)}_{M}(p_2)
+ 5 n^{(+)}_{N} (s_1)n^{(+)}_{N}(s_2)
n^{(+)}_{M}(p_1)n^{(+)}_{M}(p_2)\nonumber\\
+ 5 n^{(+)}_{N} (s_1)n^{(-)}_{N}(s_2) 
n^{(0)}_{M}(p_1)n^{(0)}_{M}(p_2)
+ 5 n^{(0)}_{N} (s_1)n^{(0)}_{N}(s_2) n^{(+)}_{M}(p_1)
n^{(-)}_{M}(p_2)\nonumber\\
+2n^{(0)}_{N}(s_1)n^{(+)}_{N}(s_2) n^{(0)}_{M}(p_1)n^{(+)}_{M}(p_2)
+ n^{(0)}_{N}(s_1)n^{(0)}_{N}(s_2)
n^{(0)}_{M}(p_1)n^{(0)}_{M}(p_2)]\nonumber\\
\times\bigg[ e^{-i(q_1-q_2)x_{NM}} 
\int_{{\cal V}_N}\!\! dy_1 e^{-i(k_1+q_1-s)y_1} 
\int_{{\cal V}_N}\!\! dy_2 e^{i(k_1+q_2-s)y_2}
\int_{{\cal V}_M}\!\! dy_3 e^{-i(k_2+q_2-p)y_3} 
\int_{{\cal V}_M}\!\! dy_4 e^{i(k_2+q_1-p)y_4}\nonumber\\ 
+e^{-i(q_1+k_1-q_2-k_2)x_{NM}}
\int_{{\cal V}_N}\!\! dy_1 e^{-i(k_1+q_1-s)y_1} 
\int_{{\cal V}_N}\!\! dy_2 e^{i(k_2+q_2-s)y_2}
\int_{{\cal V}_M}\!\! dy_3 e^{-i(k_2+q_2-p)y_3} 
\int_{{\cal V}_M}\!\! dy_4 e^{i(k_1+q_1-p)y_4}\bigg]\bigg\}~,
\label{eq:A1.1}
\end{eqnarray}
where  $s=s_1+s_2$,  $p=p_1+p_2$;
$x_N$ and $x_M$ are the ``central'' coordinates of two cells, and
$x_{NM}=x_N-x_M$. 
A quite long expression emerges for the correlator of two different pions
as well.
\begin{eqnarray}
{ dN_{\pi^{+} \pi^{-}}^{(2)} \over d\bbox k_1 d\bbox k_2}  = 
{ dN_{\pi^{+}}^{(1)} \over d\bbox k_1 } { dN_{\pi^{-}}^{(1)}\over d\bbox k_2 }
~+~{g_{4\pi}^4\over 6}\sum_{N,M} 
\int {d^4s_1 d^4s_2 d^4p_1 d^4p_2 d^4q_1 d^4q_2 
   \over 4k_1^0 k_2^0(2\pi)^{24} }\nonumber\\
\times\delta_+(q_1^2-m^2) \delta_+(q_2^2-m^2)\delta_+(s_1^2-m^2) 
\delta_+(s_2^2-m^2) \delta_+(p_1^2-m^2) \delta_+(p_2^2-m^2)  \nonumber\\
\times \bigg[ e^{-i(q_1-q_2)x_{NM}} 
[7n^{(+)}_{N} (s_1)n^{(+)}_{N}(s_2) n^{(+)}_{M}(p_1)n^{(+)}_{M}(p_2)\nonumber\\
+6 n^{(+)}_{N} (s_1)n^{(-)}_{N}(s_2) n^{(-)}_{M}(p_1)n^{(-)}_{M}(p_2)+
n^{(+)}_{N} (s_1)n^{(-)}_{N}(s_2) n^{(0)}_{M}(p_1)n^{(0)}_{M}(p_2)+
n^{(0)}_{N} (s_1)n^{(0)}_{N}(s_2) n^{(0)}_{M}(p_1)n^{(0)}_{M}(p_2)]\nonumber\\
\times\int_{{\cal V}_N}\!\! dy_1 e^{-i(k_1+q_1-s)y_1} 
\int_{{\cal V}_N}\!\! dy_2 e^{i(k_1+q_2-s)y_2}
\int_{{\cal V}_M}\!\! dy_3 e^{-i(k_2+q_2-p)y_3} 
\int_{{\cal V}_M}\!\! dy_4 e^{i(k_2+q_1-p)y_4}\nonumber\\ 
+e^{-i(q_1+k_1-q_2-k_2)x_{NM}}
[17n^{(+)}_{N}(s_1)n^{(-)}_{N}(s_2) n^{(+)}_{M}(p_1)n^{(-)}_{M}(p_2)\nonumber\\
+4n^{(+)}_{N} (s_1)n^{(-)}_{N}(s_2) n^{(0)}_{M}(p_1)n^{(0)}_{M}(p_2)
+4 n^{(0)}_{N} (s_1)n^{(0)}_{N}(s_2) n^{(+)}_{M}(p_1)n^{(-)}_{M}(p_2)
+n^{(0)}_{N} (s_1)n^{(0)}_{N}(s_2) n^{(0)}_{M}(p_1)n^{(0)}_{M}(p_2)]\nonumber\\
\times\int_{{\cal V}_N}\!\! dy_1 e^{-i(k_1+q_1-s)y_1} 
\int_{{\cal V}_N}\!\! dy_2 e^{i(k_2+q_2-s)y_2}
\int_{{\cal V}_M}\!\! dy_3 e^{-i(k_2+q_2-p)y_3} 
\int_{{\cal V}_M}\!\! dy_4 e^{i(k_1+q_1-p)y_4}\bigg]~.
\label{eq:A1.2}
\end{eqnarray} 
In Sec.~\ref{sec:SN4}, we proceed with a simplified version of these
equations which emerges when the distributions $n^{(j)}_{N}(p)$ are taken in
Boltzmann form with an additional assumption that the system is locally
neutral. This results in a common weight function and coefficients which are
just the numbers of terms in different groups in Eqs.~(\ref{eq:A1.1}) and
(\ref{eq:A1.2}).

Integration over the elementary volumes in Eqs.~(\ref{eq:A1.1}) and
(\ref{eq:A1.2}) requires special discussion: There is an important difference
between the volumes $V_N$ in equations of Sec.~\ref{sec:SN3} and the volumes
${\cal V}_N$  here. The former are related to the elementary fluid cells where
the distributions $n_{N}(p)$ of particles over their momenta are established
due to many collisions inside the $N$'th cell , and we do not (and even are
prohibited to) localize the coordinates of individual collisions within the
cell explicitly. The latter correspond to the opposite case, when we locate the
space-time coordinates of individual collisions. Therefore, the volumes
${{\cal V}_N}$ correspond to the actual range of the pion interaction
potential and are much smaller than the volumes $V_N$ of the fluid cells
required by the hydrodynamic picture. Hence, the meaning of the  distribution
functions $n_{N}(p)$ changes as well; now they correspond to the probability
for the pion with momentum ${\bbox p}$ to penetrate the ``reaction domain''
near the space-time point $x_N$. In fact, the volume ${\cal V}_N$ is defined
by the cross section of the $\pi\pi$-interaction itself. This picture is
consistent only if the pions themselves are considered as wave packets and
not as plane waves. The volume ${\cal V}_N$ is that volume where the incoming
packets are identified by the interaction and where the outgoing packets are
completely formed.  With this picture in mind, we can integrate, e.g.,
\begin{eqnarray}
\int_{{\cal V}_N}\!\! dy_1 e^{-i(k_1+q_1-s)y_1} 
\int_{{\cal V}_N}\!\! dy_2 e^{i(k_1+q_2-s)y_2}=
(2\pi)^4\delta(k_1+q_1-s)\int_{{\cal V}_N}\!\! dy_2 e^{-i(q_1-q_2)y_2}
= (2\pi)^4\delta(k_1+q_1-s){\cal V}_N~.
\label{eq:A1.3}
\end{eqnarray}
In the first equation, we assume that the relation between the volume of
integration and momenta allows one to verify the conservation of momentum
in the collision. In the  second equation, we took into account that the actual
range for the momenta  $q_1$ and $q_2$ is of the order of $2T\sim 2m_\pi$,
while the length of the $\pi$-$\pi$ scattering is $\sim 1/5m_\pi$. In all
subsequent calculations we can rescale ${\cal V}_N$ back to the elementary 
volume $V_N$ and thus restore the status of $n_N(p)$ as the distribution 
function.

\renewcommand{\theequation}{B.\arabic{equation}}
\setcounter{equation}{0}

\section*{Appendix B. Miscellaneous formulae.}
\label{sec:A2}

1.~ If we adopt the Boltzmann shape of the distribution function, then 
the product of the statistical weights  $n(p_1)n(p_2)$ becomes a function of
$p=p_1+p_2$ and it is possible to integrate $l=p_1-p_2$ out. We have
an integral,
\begin{eqnarray}
\int d^4p_1 d^4p_2 f(p) \delta_+(p_1^2-m^2) \delta_+(p_2^2-m^2)=
{1\over 8}\int d^4p  f(p) \int d^4l \delta(p^2+l^2-4m^2)\delta(pl)\nonumber\\
=\int d^4p  f(p){\pi\over 2p^0}\int_{0}^{\infty}
\delta(p^2-4m^2+{\bbox l}^2)~{\bbox l}^2~d |{\bbox l}|=
{\pi \over 4}\int d^4p  f(p) \sqrt{1-{4m^2\over p^2}}~,
\label{eq:A2.0}
\end{eqnarray}
where in the second equation we integrate using a special reference frame where
${\bbox p}=0$, and restore the invariant form in the final answer.

2.~ The correlation functions of Sec.~\ref{sec:SN4} are expressed via
the emission function $I(k,N,x_{NM})$ which is computed below:
\begin{eqnarray}
I(k,N,x_{NM})=\int d^4q\delta_+(q^2-m^2)e^{-\beta(qu)}~e^{iqx}
\sqrt{1-{4m^2\over (k+q)^2}}~,
\label{eq:A2.1}
\end{eqnarray}
where $\beta=\beta_N$, $u=u_N$, and $x=x_{NM}$. The integral is convenient to
compute in a special reference frame  $ \overcirc{\cal R}$ where the time-like
vector $k$ has no spatial components, ${\overcirc{\bbox k}}=0$, ${\overcirc
k}_0=m$. Using the delta-function to integrate out $q^0$, we arrive at 
\begin{eqnarray}
I(k,N,x_{NM})=\int_{0}^{2\pi}d\phi\int_{0}^{\pi}\sin\alpha d\alpha
\int_{0}^{\infty}{qdq\over 2\sqrt{q^2+m^2}}\nonumber\\
\times\exp\{-\sqrt{q^2+m^2}(\beta -i{\overcirc x}_0)+
q[\beta |{\overcirc{\bbox u}}|\cos\alpha-i
|{\overcirc{\bbox x}}|(\cos\alpha\cos\psi -\sin\alpha\sin\psi\cos\phi)]\}~.
\label{eq:A2.2}
\end{eqnarray} 
The first step is to integrate over $\phi$ which results in the Bessel function,
$2\pi J_0(|{\overcirc{\bbox x}}|q\sin\alpha\sin\psi)$. The next integration
follows a known formula,
\begin{eqnarray}
\int_{0}^{\pi}d\alpha~ \sin\alpha~  e^{ib\cos\alpha} J_0(a\sin\alpha)=
2{\sin\sqrt{a^2+b^2}\over\sqrt{a^2+b^2}}~,
\label{eq:A2.3}
\end{eqnarray}
which holds for arbitrary complex $a$ and $b$. In our case, $a^2+b^2=-(\beta
{\overcirc{\bbox u}}-i{\overcirc{\bbox x}})^2=-{\overcirc{\bbox U}^2}$
where we have introduced the complex four-vector $U^\mu=\beta u^\mu - i x^\mu$.
Finally, changing the variable of integration, $q=m\sinh v$, we arrive at
\begin{eqnarray}
I(k,N,x_{NM})={2\pi m\over|\overcirc{\bbox U}|}
\int_{0}^{\infty}(\cosh v -1) dv~
e^{-m {\overcirc U}_0 \cosh v}
\sinh(m |\overcirc{\bbox U}|\sinh v)\nonumber\\
={2\pi m\over \sqrt{U^2}\sinh\chi}
\bigg[\int_{0}^{\infty}dv~
(\cosh\chi \cosh v -1)e^{-m \sqrt{U^2} \cosh v}-
\int_{0}^{\infty}dv~(\cosh v -1) e^{-m \sqrt{U^2} \cosh (v+\chi)}\bigg]~.
\label{eq:A2.4}
\end{eqnarray} 
The Lorentz-invariant expressions for the quantities defined in the reference
frame $ \overcirc{\cal R}$ are
\begin{eqnarray}
m{\overcirc U}_0 = (kU)=m\sqrt{U^2} \cosh\chi,~~
m|\overcirc{\bbox U}| = \sqrt{(kU)^2-m^2U^2}=m\sqrt{U^2}\sinh\chi~.
\label{eq:A2.5}
\end{eqnarray} 
The first of the integrals in Eq.~(\ref{eq:A2.4}) is calculated exactly.
The second one can be estimated  by means of integration by parts. Since the
integrand has the root of the second order
 at $v=0$, we have to integrate by parts three
times before the first non-vanishing term shows up. The net yield of this
procedure is
\begin{eqnarray}
-{4\pi m\over |\overcirc{\bbox U}|}\int_{0}^{\infty}dv~\sinh^2{ v \over 2} 
e^{-m \sqrt{U^2} \cosh (v+\chi)}= 
-{4\pi m^2\over m^4|\overcirc{\bbox U}|^4}e^{-m {\overcirc U}_0}
+\cdot\cdot\cdot~.
\label{eq:A2.6}
\end{eqnarray} 
Since $m {\overcirc U}_0= (kU)\equiv (ku)/T -i(kx_{NM})$, 
 this term is strongly 
suppressed (both by the exponent and by at least the fourth power of the small 
number $m/k_t$). Finally, we have to high accuracy,
\begin{eqnarray}
I(k,N,x_{NM})={2\pi m\over \sqrt{U^2}}\bigg[\coth\chi K_1(m\sqrt{U^2})-
{1\over \sinh\chi}K_0(m\sqrt{U^2})\bigg]~.
\label{eq:A2.7}
\end{eqnarray}
Here, $K_n(z)$ is the standard notation for the modified Bessel function, and 
we remind the reader that $U^2$ and $\chi$ (as well as ${\overcirc U}_0$ and 
$|\overcirc{\bbox U}|$) all are complex quantities. When $x\equiv x_{NM}=0$,
we have $U^2=\beta^2$, $\cosh\chi=(ku)/m=(m_t/m)\cosh (\theta -\eta)\gg 1$,
and consequently, $\sinh\chi\gg 1$, $\coth\chi\sim 1$. Therefore,
\begin{eqnarray}
I(k,N,0)=2\pi m^2 {K_1(\beta m)\over \beta m}~.
\label{eq:A2.8}
\end{eqnarray}
In  the particular cases considered in Sec.~\ref{sec:SN4}, the second term in
Eq.~(\ref{eq:A2.7}) is always smaller than the first one by the factor $m/k_t$.
An important property of the final answer (\ref{eq:A2.7}) is that
the quantity
\begin{eqnarray}
U^2=\beta^2(x_1) -(x_1-x_2)^2-2i\beta (x_1)(u(x_1)\cdot(x_1-x_2))\nonumber
\end{eqnarray}
in the argument of function $K_1$, does not depend on the large parameter 
$k_t/m$. Therefore, the saddle point of the integration over the coordinate
rapidities $\eta$ is never affected by this function.

\end{document}